\documentclass[aps,prb,10pt%,preprint%,longbibliography
]{revtex4-2}
\usepackage{graphicx,bm,color,%comment,
amssymb}

\def\be{\begin{equation}}
\def\ee{\end{equation}}
\def\ba{\begin{eqnarray}}
\def\ea{\end{eqnarray}}

\def\bc{\begin{center}}
\def\ec{\end{center}}

\begin{document}

\title{Toward a new theory of the fractional quantum Hall effect: \\ The many-body spectra and energy gaps at $\nu<1$}
\author{S. A. Mikhailov}%\cite{email}}
\email[Electronic mail: ]{sergey.mikhailov@physik.uni-augsburg.de} 
\affiliation{Institute of Physics, University of Augsburg, D-86135 Augsburg, Germany} 

\date{\today}

\begin{abstract}
In a recent paper (arXiv:2206.05152v4), using the exact diagonalization technique, I calculated the energy and other physical properties (electron density, pair correlation function) of a system of $N\le 7$ two-dimensional electrons at the Landau level filling factor $\nu=1/3$, and showed that the variational many-body wave function proposed for this filling factor by Laughlin is far from the true ground state. In this paper I continue to study exact properties of a small ($N\le 7$) system of two-dimensional electrons lying on the lowest Landau level. I analyze the energies and electron densities of the systems with $N\le 7$ electrons continuously as a function of the magnetic field in the range $1/4\lesssim\nu<1$. The physical mechanisms of the appearance of energy gaps in many-particle electron spectra are elucidated. The results obtained clarify the true nature of the ground and excited states of the considered systems.
\end{abstract}

\maketitle

\tableofcontents

\section{Introduction\label{sec:intro}}

The fractional quantum Hall effect (FQHE) was experimentally discovered in 1982 \cite{Tsui82}. The authors measured the Hall ($\rho_{xy}$) and diagonal resistivities ($\rho_{xx}$) of a high-mobility two-dimensional (2D) electron gas in a GaAs/AlGaAs heterostructure, placed in a strong magnetic field $B$ (up to $\sim 22$ T) and cooled down to very low temperatures ($T\sim 0.48$ K). It was found that, when the Landau level filling factor 
\be 
\nu=2\pi n_s\frac{\hbar c}{eB}=n_s\frac{h c}{eB}\label{LLff}
\ee
is close to $\nu\approx 1/3$, the Hall resistivity $\rho_{xy}$ of the system is ``quantized'',
\be 
\rho_{xy}= 3\frac h{e^2},\label{fracRxy}
\ee
while the diagonal resistivity $\rho_{xx}$ has a deep minimum; here $n_s$ is the 2D electron gas density and $h=2\pi\hbar$ is the Planck constant. The observed phenomenon looked very similar to the \textit{integer} quantization of the Hall resistivity, 
\be 
\rho_{xy}= \frac h{e^2i},\ i=1,2,\dots,\label{integerRxy}
\ee
discovered earlier \cite{Klitzing80} in the vicinity of integer values of $\nu\approx i$. However, while the physical reason of the integer quantum Hall effect (IQHE) was quite quickly understood, the physics of the fractional quantization looked rather mysterious. Indeed, the spectrum of 2D electrons in a perpendicular magnetic field is quantized \cite{Landau30},
\be 
E_n=\hbar\omega_c\left(i+\frac 12\right), \ i=0, 1,\dots,
\ee
so that at integer values of $\nu$ there exist large energy gaps in the single-particle spectrum of electrons; here $\omega_c=eB/m^\star c$ is the cyclotron frequency, and $m^\star$ is the effective mass of 2D electrons . Substituting the density of electrons from (\ref{LLff}) into the classical Drude formula for the Hall resistivity $\rho_{xy}=B/n_sec$, one immediately gets the quantized values (\ref{integerRxy}) when $\nu=i$ is integer; the stabilization of $\rho_{xy}$ at the levels  (\ref{integerRxy}) and the vanishing of $\rho_{xx}$ in finite intervals around $\nu=i$ was explained by the influence of disorder, see, e.g., Ref. \cite{Prange90}. At $\nu<1$, in contrast, the Landau level is highly degenerate, and there is no gap in the single-particle electron spectrum. It became clear that in order to explain the fractional quantization (\ref{fracRxy}) at $\nu=1/3$, the single-particle approach is not enough, and a many-body theory is required.

In 1983 Laughlin \cite{Laughlin83} proposed a trial wave function 
\be 
\Psi_{\rm RL}^{m}(\bm r_1,\bm r_2,\dots,\bm r_N)\propto \left(\prod_{1\le j<k\le N} (z_j-z_k)^m \right)\exp\left(-\frac 12\sum_{j=1}^N|z_j|^2\right)
\label{LaughlinWF}
\ee
for the ground state of an $N$-particle system at $\nu=1/m$. Here $m=3,5,\dots$ is odd integer, $\bm r_j=(x_j,y_j)$, and $z_j=(x_j-iy_j)/\lambda$ are the normalized complex coordinates of 2D electrons, where 
\be 
\lambda=\sqrt{2}l_B=\sqrt{\frac{2\hbar c}{|e|B}}=\sqrt{\frac{2\hbar}{m^\star\omega_c}},\label{lambda}
\ee 
and $l_B$ is the magnetic length. If $m=1$, the function (\ref{LaughlinWF}) coincides with the wave function of the so called maximum density droplet (MDD) state proposed earlier in Ref. \cite{Bychkov81} for the ground state of the system at $\nu=1$. For $m=3$ and 5 Laughlin calculated the energy (per particle) of the state (\ref{LaughlinWF}) in the thermodynamic limit,
\be 
\frac{E_{\rm RL}^{m=3}}{N(e^2/l_B)}=-0.4156\pm 0.0012,\ \ 
\frac{E_{\rm RL}^{m=5}}{N(e^2/l_B)}=-0.3340\pm 0.0028, \label{LaughEnergyPP}
\ee 
and found that these values are lower than the corresponding numbers for the charge-density wave \cite{Yoshioka79,Yoshioka83b}. He also ``generated'' elementary excitations of $\Psi_{\rm RL}^{m}$ by multiplying it by additional polynomial factors $\prod_i(z_i-z_0)$ and claimed that these excitations are particles of charge $e/m$. These statements have not been mathematically proved.

Subsequent experimental studies showed that the fractional quantization of the Hall resistivity occurs around many fractions of the form $\nu=p/q$ where $p$ and $q$ are integers and $q$ is odd, as well as around some fractions with an even denominator \cite{Willett87}. Various theoretical approaches have been proposed to explain the new fractions, for example, hierarchical schemes \cite{Haldane83} or the composite fermions theory \cite{Jain89}. In the latter theory, the FQHE of electrons is treated as the IQHE of composite fermionic objects consisting of electrons bound to an even number of flux quanta. %It is also rather unclear, how such really existing particles as electrons attach to themselves such mathematical objects as magnetic flux quanta.

In a recent paper \cite{Mikhailov23a} I developed an exact diagonalization theory that makes it possible to calculate the energy and other physical properties (electron density, pair-correlation function) of the ground and excited states of a system of $N$ two-dimensional Coulomb interacting electrons in a strong magnetic field. I have studied physical properties of the exact ground state for a system of $N\le 7$ electrons at $\nu=1/3$, and found that it resembles the floating Wigner crystal: the density of electrons has the shape of a ring of radius $\approx R_s$, with an additional maximum at $r=0$ if $N\ge 6$, where $R_s$ is the radius of the shells in the classical $N$-electron Wigner molecules. I have also compared the properties of the ``Laughlin liquid'' (\ref{LaughlinWF}) with those of the exact ground state and found a very large and growing with $N$ discrepancy between them. In particular, I have shown that at $N=7$
\begin{itemize}
\item the energy difference between the Laughlin and the true ground state is almost six times larger than the energy difference between the true first excited and the true ground state;
\item the electron density difference reaches $\sim 70$\% in the center of a disk-shaped sample;
\item the pair correlation functions $P(\bm r,\bm r')$ of the Laughlin and the true ground state differ by many orders of magnitude if $|\bm r-\bm r'|\lesssim 0.3a_0$, where $a_0$ is the average distance between 2D electrons.
\end{itemize}

I have also analyzed the ``thermodynamic-limit argument'' of Ref. \cite{Laughlin83}, that the function (\ref{LaughlinWF}) gives the lowest possible energy per particle (\ref{LaughEnergyPP}) in the limit $N\to\infty$. I have proved that, see Sections I.B and V.H of Ref. \cite{Mikhailov23a}, while the variational principle may give a correct estimate for the ground state energy per particle, it completely fails to determine the correct ground state wave function. One can specify an infinite number of trial wave functions with completely unphysical properties, which, however, will have the same energy per particle in the thermodynamic limit as the true ground state.

In this paper, I continue to develop the theory of Ref. \cite{Mikhailov23a} and analyze the exact properties of $N\le 7$ 2D electrons placed in the field of a positively charged background having the form of a disk. I study the system properties in the range $1/4\lesssim\nu<1$ as a function of the magnetic field. I show that both the ground and excited states of the system have the shapes resembling the floating Wigner molecules: if $N\le 5$, the density of electrons has the form of a ring with a maximum at $r=R_{\max}$, where $R_{\max}$ is close to the shell radius $R_s$ of a classical Wigner molecule. If $N\ge 6$ an additional density maximum appears at $r=0$, in full agreement with the classical distribution of point charges in the field of the positively charged disk. Parameters of the density rings -- their radius, width and the density value in the maximum -- depend on magnetic field. When $B$ varies, the ring radius $R_{\max}$ oscillates around $R_s$. The distance between the first excited state and the ground state, that is the energy gap in the  many-particle spectra of the system, oscillates as a function of $B$, vanishing at certain points of the $B$ axis and remaining finite in finite-width intervals of $\nu$. The absolute values of the energy gap are on the fractions-of-meV scale for parameters of typical semiconductor FQHE structures.

The rest of the paper is organized as follows. In Section \ref{sec:theory} I formulate the problem, derive missing formulas necessary to calculate the energy spectra and other physical properties of the system at arbitrary magnetic fields, and give a short overview of the exact diagonalization procedure. Section \ref{sec:results} contains the results of the work for the energy (Section \ref{sec:ResSpectra}) and the electron density (Section \ref{sec:ResDensity}). In addition to the results for the step-like density profile of a positively charged background disk \cite{Mikhailov23a}, I also present here some results for a smooth density profile which gives a more adequate description of real systems. Section \ref{sec:conclusion} contains a summary of results and conclusions. Some details of the calculations are given in the Appendix \ref{app:intJ}.

\section{Theory\label{sec:theory}}

\subsection{Formulation of the problem\label{sec:formulation}}

I consider $N$ 2D electrons moving in the plane $z=0$ in the field of a positively charged background and in the presence of a uniform external magnetic field $\bm B=(0,0,B)$. The positive background has the shape of a disk of radius $R$ and the background charge density $+|e|n_s$. All lengths and energies in the paper will be measured in units $a_0$ and $e^2/a_0$, where $a_0$ is the average distance between electrons defined as 
\be 
\pi n_sa_0^2=1. \label{def-a0}
\ee
I do not explicitly write the dielectric permittivity of the environment $\epsilon$ in formulas containing Coulomb forces and energies; to restore it, one should replace $e^2\to e^2/\epsilon$.

To find the energies and many-body wave functions of the system one should solve the $N$-particle Schr\"odinger equation
\be 
\hat {\cal H}|\Psi\rangle=E|\Psi\rangle,\ \ \ |\Psi\rangle=\Psi(\bm r_1,\bm r_2,\dots,\bm r_N),
\label{NbodySchrEq}
\ee
with the Hamiltonian 
\be 
\hat {\cal H}=\hat K+\hat V_C =\frac{1}{2m^\star}\sum_{j=1}^N\left(\hat {\bm p}_j+\frac {|e|}{2c}{\bm B}\times{\bm
r}_j\right)^2 +\hat V_C ,
\label{MBhamilt}
\ee
where the vectors $\hat {\bm p}$ and $\bm r$ are two-dimensional, $\hat K$ is the total kinetic energy operator and $\hat V_C=\hat V_{bb}+\hat V_{eb}+\hat V_{ee}$ is the Coulomb interaction energy. The latter consists of the sum of background-background, background-electron and electron-electron interaction energies, 
\ba 
\hat V_C
&=&\frac{e^2}{2}\int \frac{n_b(\bm r)n_b(\bm r')d\bm r d\bm r'}{|\bm r-\bm r'|}
-e^2 \int n_b(\bm r) d\bm r\sum_{j=1}^N \frac 1{|\bm r-\bm r_j|}
+\frac{e^2}{2}\sum_{j\neq k=1}^N \frac 1{|\bm r_j-\bm r_k|}
\nonumber \\
&=&\frac{e^2}{2\pi} \int \frac{d\bm q}{q}
\left(
\frac{1}{2}\left(n_{\bm q}^b\right)^2 
- n_{\bm q}^b \sum_{j=1}^N  e^{-i\bm q\cdot \bm r_j}
+\frac{1}{2}\sum_{j\neq k=1}^N e^{i\bm q\cdot \bm r_j} e^{-i\bm q\cdot \bm r_k}
\right),
\label{CoulombEnergy}
\ea
where $n_b(\bm r)$ and $n_{\bm q}^b$ are the density of the positively charged background and its Fourier transform. The Hamiltonian (\ref{MBhamilt}) commutes with the total angular momentum operator 
\be 
\hat {\cal L}_z=\sum_{j=1}^N({\bm r}_j \times \hat {\bm p}_j)_z.
\ee
The eigenvalues of this operator, 
\be 
\hat {\cal L}_z|\Psi\rangle={\cal L}|\Psi\rangle,\label{LtotDefinition}
\ee 
the total angular momentum quantum number ${\cal L}$, can be used for classifying the many-body eigenstates of the Hamiltonian (\ref{MBhamilt}).

In Ref. \cite{Mikhailov23a} I considered the model of a step-like density profile, for which 
\be 
n_b^{\rm st}(\bm r)=n_s\Theta(R-r)=n_s\Theta(\sqrt{N}-x),
\label{dens-step}
\ee 
where $R=\sqrt{N/\pi n_s}$, $\Theta(x)$ is the Heaviside step function, and  $x=r/a_0$. In this paper, along with the model (\ref{dens-step}), I will also consider the model of a smooth density profile
\be
n_b^{\rm sm}({\bm r})= n_se^{-x^2}\sum_{k=0}^{N-1}\frac{(x^2)^k}{k!}=n_s\frac{\Gamma(N,x^2)}{\Gamma(N)}=n_s Q(N,x^2),
\label{dens-smooth}
\ee
where $\Gamma(N)$ is the Euler Gamma function, $\Gamma(N,z)$ is the incomplete Gamma function, and
\be 
Q(N,x)=\frac{\Gamma(N,x)}{\Gamma(N)} \label{Qfunc}
\ee 
is the regularized incomplete Gamma function. The smooth density profile (\ref{dens-smooth}) actually gives a more correct description of the real density distribution, since in real systems the edge is always smeared over a certain length, for example, at the distance between the 2D gas and the donor layer in GaAs/AlGaAs heterostructures or at the average distance between electrons $a_0 $.

\begin{figure}[ht!]
\includegraphics[width=0.5\columnwidth]{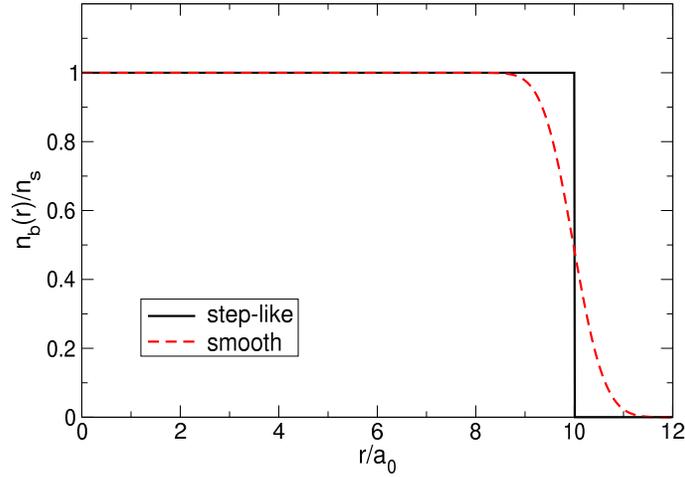}
\caption{\label{fig:bcgrdens}The positive background density $n_b (r)$ for $N = 100$ in the smooth and step-like profile models.
}
\end{figure}

The both density profiles, (\ref{dens-step}) and (\ref{dens-smooth}), satisfy the condition 
\be 
\int n_b({\bm r}) d\bm r =N
\ee
and are shown in Figure \ref{fig:bcgrdens} for $N=100$. The Fourier transforms of the density profiles (\ref{dens-step}) and (\ref{dens-smooth}) are determined by the formulas
\be
n_{\bm q}^{b,\rm st}\equiv\int d\bm r n_b({\bm r})e^{i\bm q\cdot\bm r}=N \frac {J_1(qR)}{qR/2}=2N \frac {J_1(Q\sqrt{N})}{Q\sqrt{N}},
\label{BGdensStep_Fourier}
\ee
and
\be 
n_{\bm q}^{b,\rm sm}=e^{-Z}
 L_{N-1}^1\left(Z\right) ,\label{BGdensSmooth_Fourier}
\ee
respectively, where $Q=qa_0$,  $Z=Q^2/4$, $J_1$ is the Bessel function, and $L_n^k(Z)$ are the Laguerre polynomials. 

\begin{figure}[ht!]
\includegraphics[width=0.49\columnwidth]{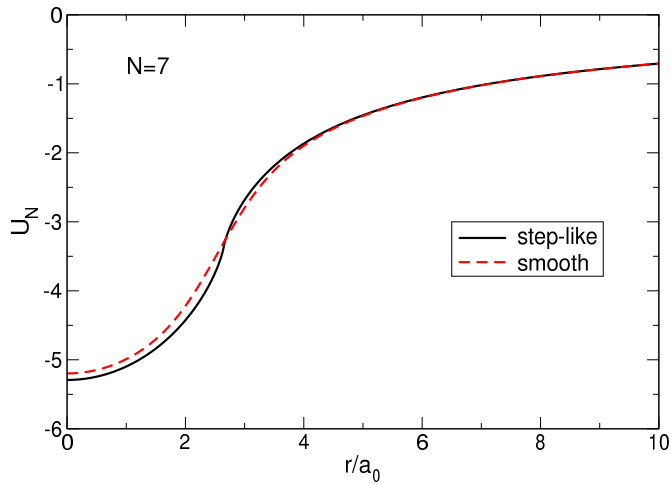}
\caption{\label{fig:Ubpotential}The potential energies (\ref{Upot-step}) and (\ref{Upot-smooth}) of the positive background with the density profiles (\ref{dens-step}) and (\ref{dens-smooth}) for $N=7$.}
\end{figure}

The potential well created by the positively charged background disk with the density (\ref{dens-step}) or (\ref{dens-smooth}) is described by the formula
\be 
V_b(\bm r)=\frac{e^2}{a_0}U_N\left(\frac{r}{a_0}\right),\label{Upot}
\ee
where
\be 
U_N^{\rm st}\left(x\right)=
-\frac 4\pi \left\{
\begin{array}{ll}
\sqrt{N}\textrm{E}\left(\frac{x^2}{N}\right), & x^2\le N \\
x\textrm{E}\left(\frac{N}{x^2}\right)-\frac{x^2-N}{x}\textrm{K}\left(\frac{N}{x^2}\right), & x^2\ge N \\
\end{array}
\right. ,
\label{Upot-step}
\ee
for the step-like density profile (\ref{dens-step}), and 
\be 
U_N^{\rm sm}(x)=-\sum_{m=0}^{N-1}
\left(
\begin{array}{c}
N \\ m+1 \\
\end{array}
\right)
\frac{(-1)^m }{m!} \Gamma\left(m+\frac 12\right){_1F_1}\left(m+\frac 12,1;-x^2\right) \label{Upot-smooth}
\ee
for the smooth density profile (\ref{dens-smooth}). Here the functions $\textrm{K}(m)$ and $\textrm{E}(m)$, defined as 
\be  
\textrm{K}(m)=\int_0^{\pi/2}\frac{d\theta}{\sqrt{1-m\sin^2\theta}},\ \ \ \textrm{E}(m)=\int_0^{\pi/2}d\theta\sqrt{1-m\sin^2\theta},
\ee 
are complete elliptic integrals of the first and second kind, respectively, $\left(\begin{array}{c}
n \\ m \\
\end{array}\right)$ are the binomial coefficients and ${_1F_1}\left(a,b;z\right)$ is the degenerate (confluent) hypergeometric function
\be 
{_1F_1}\left(a,b;z\right)=\sum_{k=0}^\infty \frac{(a)_k}{(b)_k}\frac{z^k}{k!};
\label{DegenHyperGfun}
\ee 
$(a)_k=\Gamma(a+k)/\Gamma(a)$ is the Pochhammer symbol. Figure \ref{fig:Ubpotential} shows the functions (\ref{Upot-step}) and (\ref{Upot-smooth}) for $N=7$. For the step-like density profile, the potential well is deeper, and its effective width is slightly bigger. The depths of both potential wells grow with $N$ as $\sqrt{N}$, $U_N^{\rm sm}(0)\approx U_N^{\rm st}(0)=-2\sqrt{N}$.

\subsection{The exact diagonalization procedure\label{sec:EDprocedure}}

The many-body states of the Hamiltonian (\ref{MBhamilt}) can be classified by the quantum number ${\cal L}$, Eq. (\ref{LtotDefinition}). For any value of the particle number $N$ I perform calculations for ${\cal L}$ lying in the interval ${\cal L}_{\min}\le {\cal L}\le {\cal L}_{\max}$, where 
\be 
{\cal L}_{\min}=\frac{N(N-1)}{2},\ \ {\cal L}_{\max}=4{\cal L}_{\min}.
\label{MinMaxMomenta}
\ee
I assume that all electrons are fully spin-polarized and occupy only the lowest Landau level single-particle states
\be
|L\rangle\equiv \psi_L({\bm r})= \frac{z^{L} e^{-|z|^2/2}}{\lambda\sqrt{\pi L!}} ,
\label{spwf_psiL}
\ee
where $L=0,1,2,\dots $, and $z=(x-iy)/\lambda=re^{-i\phi}/\lambda$; $\phi$ is the polar angle. For every pair of numbers $N$ and ${\cal L}$, I calculate the number of many-body states $N_{mbs}=N_{mbs}(N,{\cal L})$ and expand the function $|\Psi\rangle$ in (\ref{NbodySchrEq}) in a set of Slater determinants corresponding to the given values of $N$ and ${\cal L}$, 
\be 
|\Psi\rangle =\sum_{s'=1}^{N_{mbs}} A_{s'} |\Psi_{s'}\rangle.
\label{PsiExpansion}
\ee
Here
\be 
|\Psi_{s}\rangle=|L_1^s,L_2^s,\dots,L_N^s\rangle=
\frac{1}{\sqrt{N!}}
\left |\begin{array}{cccc}
\psi_{L_1^s}(\bm r_1) & 
\psi_{L_1^s}(\bm r_2) & 
\dots &
\psi_{L_1^s}(\bm r_N) \\
\psi_{L_2^s}(\bm r_1) & 
\psi_{L_2^s}(\bm r_2) &
\dots & 
\psi_{L_2^s}(\bm r_N) \\
\dots &\dots &\dots &\dots \\
\psi_{L_N^s}(\bm r_1) & 
\psi_{L_N^s}(\bm r_2) & 
\dots & 
\psi_{L_N^s}(\bm r_N) \\
\end{array}
\right |
\label{manybodyWF}
\ee
and $L_1^s+L_2^s+\dots+L_N^s={\cal L}$. Then the problem is reduced to the eigenvalue problem 
\be 
\sum_{s'=1}^{N_{mbs}} {\cal H}_{ss'}A_{s'}=EA_{s}.
\label{matrix-equation}
\ee
for the matrix ${\cal H}_{ss'}\equiv \langle \Psi_{s}|\hat {\cal H}|\Psi_{s'}\rangle$ of the size $N_{mbs}\times N_{mbs}$. Solving the problem (\ref{matrix-equation}) I get, for given $N$ and ${\cal L}$, $N_{mbs}$ many-body energy levels $E_{N,{\cal L},k}$ and sets of numbers $A_{s}^{N,{\cal L},k}$, $k=1,\dots,N_{mbs}$, which give the corresponding many-body wave functions according to the expansion (\ref{PsiExpansion}). After the numbers $A_{s}^{N,{\cal L},k}$ are found one can calculate all physical properties of the ground and excited many-body states, for example, the electron density.

For the step-like density profile the matrix elements of the Hamiltonian ${\cal H}_{ss'}$ have been analytically calculated in Ref. \cite{Mikhailov23a}. For the smooth density profile the matrix elements of the background-background and background-electron interactions are additionally required. Calculations give
\be 
\langle \Psi_s|\hat V_{bb}^{\rm sm}|\Psi_{s'}\rangle =\delta_{ss'}
\frac{e^2}{a_0} \sqrt{\frac\pi 8}{\cal J}(N-1,N-1,1,1,0;1,1),
\label{BB_smooth_matrixelements}
\ee
\be 
\langle \Psi_{s}| \hat V_{be}^{\rm sm}|\Psi_{s'}\rangle=
-\delta_{ss'} 
\frac{e^2}{a_0} \sqrt{\frac{\pi\beta} 2}
\sum_{j=1}^N {\cal J}(L_j^{(s)},N-1,0,1,0;1,\beta).
\label{BE_smooth_matrixelements}
\ee 
where the integral ${\cal J}(n_1,n_2,l_1,l_2,k;\alpha,\beta)$ is defined and calculated in Appendix \ref{app:intJ}, see Eqs. (\ref{intJ}) and (\ref{intJsolution}), and 
\be 
\beta=\frac{a_0^2}{\lambda^2}=\frac 1\nu
\ee 
is the inverse Landau level filling factor. The matrix elements of the electron-electron interaction are determined by the same formulas as for the step-like profile and can be found in Ref. \cite{Mikhailov23a}, see Section II.E.5 there.

In the next Section I present results for the energies of the ground and several excited states, as well as for the corresponding electron densities, for the number of particles varying from $N=2$ up to $N=7$, in various magnetic fields in the range $1\le \beta\le 4$ (for $N=7$ in the range $1\le \beta\lesssim 3.5$). It should be noted that, \textit{within the framework of the assumptions made} (all electrons are spin polarized and occupy only the lowest Landau level) the results can be obtained with very high accuracy, since the problem is reduced to the diagonalization of a finite-size Hamiltonian matrix, all elements of which are calculated analytically. However, \textit{by themselves, these assumptions become insufficient} for obtaining high-accuracy solutions as the filling factor approaches $\nu=1$. Obviously, the accuracy of the results depends on the number of the basis functions $N_{mbs}$ in which the solution is expanded: the larger $N_{mbs}$, the higher the accuracy. However, when $\nu$ is close to 1 (for example, for $N=7$ at $1\le \beta\lesssim 1.4$), the ground state of the system, calculated within the approximations made, is the MDD state for which $N_{mbs}=1$. In order to get more accurate results for these $\nu$'s (close to 1), one should take into account single-particle electron states from the higher Landau levels. This, more general problem, will not be addressed in the present  paper.

\section{Results\label{sec:results}}

\subsection{Energy spectra vs. magnetic field\label{sec:ResSpectra}}

In this Section I study the magnetic field dependencies of energies of the many-body quantum states with different total angular momenta ${\cal L}$ from the interval ${\cal L}_{\min}\le {\cal L}\le {\cal L}_{\max}$, Eq. (\ref{MinMaxMomenta}), for systems of $2\le N\le 7$ particles. The ground and the first excited states are identified in different intervals of the magnetic field. The positive background density profile is assumed to be step-like everywhere except the Sections \ref{sec:resultsSmooth} and \ref{sec:density7smoo}. 

\subsubsection{Two particles}

If $N=2$, the relevant total angular momenta (\ref{MinMaxMomenta}) vary from ${\cal L}_{\min}=1$ up to ${\cal L}_{\max}=4$. The many-body configurations for these ${\cal L}$ and the number of these configurations $N_{mbs}$ are listed in Table \ref{tab:N2configs}. If (for any $N$) ${\cal L}={\cal L}_{\min}$ (the MDD configuration) or ${\cal L}={\cal L}_{\min}+1$, there exists only one many-body state, $N_{mbs}=1$, and there exists only one energy-vs-$B$ curve $E_{\cal L}(\beta)$. If ${\cal L}_{\min}>{\cal L}_{\min}+1$, the number of many-body state is bigger than one, $N_{mbs}>1$, and the energy-vs-$B$ curves $E_{{\cal L},s}(\beta)$ will be enumerated by an additional index $s$. The goal is to determine, for each $\beta$-value, the energies of the ground $E_{\rm GS}(\beta)$ and the first excited $E_{\rm 1st}(\beta)$ states, as well as the values of the quantum numbers $({\cal L},s)$ corresponding to these states.

\begin{table}[!ht]
\caption{Many-body configurations with ${\cal L}_{\min}\le {\cal L}\le {\cal L}_{\max}$ in a system of $N=2$ electrons. $N_{mbs}$ is the total number of all many-particle configurations with the given $N$ and ${\cal L}$. \label{tab:N2configs}}
\begin{tabular}{clc}
${\cal L}$ & Configurations  & $N_{mbs}$\\
\hline
1 & \hspace{3mm} $|0,1\rangle $ & 1 \\
2 & \hspace{3mm} $|0,2\rangle $ & 1 \\
3 & \hspace{3mm} $|0,3\rangle $ $|1,2\rangle $ & 2 \\
4 & \hspace{3mm} $|0,4\rangle $ $|1,3\rangle $ & 2 \\
\end{tabular}
\end{table}

\begin{figure}[ht!]
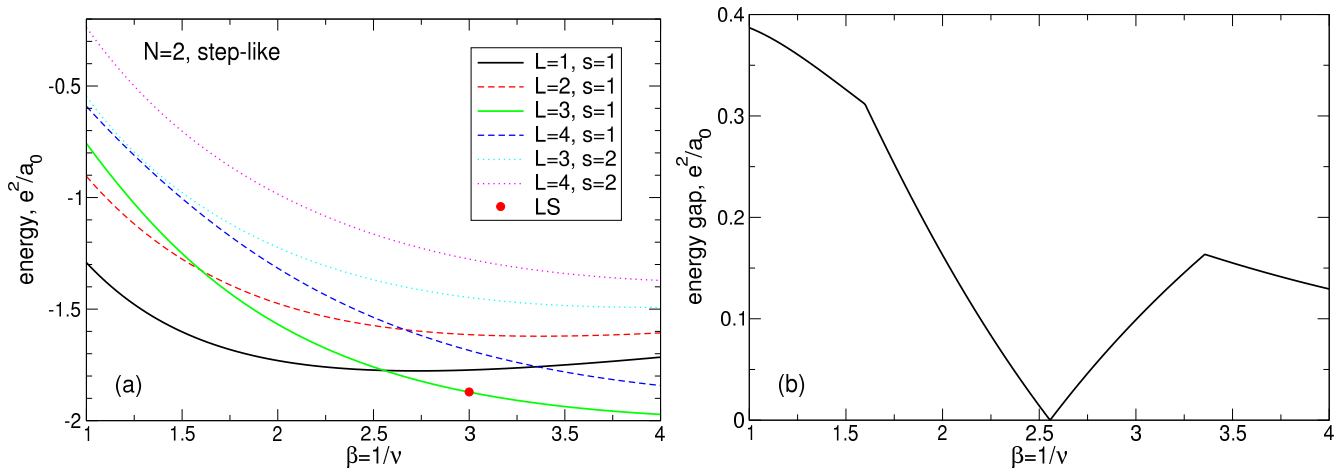

\includegraphics[width=0.49\columnwidth]{N2step.eps}
\includegraphics[width=0.49\columnwidth]{N2stepGap.eps}
\caption{\label{fig:energyN2step} (a) The energy of the many-body states with the total angular momenta from ${\cal L}={\cal L}_{\min}=1$ up to ${\cal L}={\cal L}_{\max}=4$ in the system of $N=2$ 2D electrons as a function of the magnetic field parameter $\beta=1/\nu$. The index $s$ enumerates different states with the same ${\cal L}$. The energy of the Laughlin state (LS) at $\nu=1/3$ is shown by a small red circle. (b) The energy gap between the ground and the first excited states as a function of $\beta$. The positive background density profile is step-like. }
\end{figure}

Figure \ref{fig:energyN2step}(a) shows the energies of all many-body states, with ${\cal L}$ varying from 1 to 4 and $s=1,2$, for the system of $N=2$ electrons. If $\beta=1/\nu=1$, the ground state is characterized by the quantum numbers $({\cal L},s)=(1,1)$ and is the MDD state $|0,1\rangle$. As $\beta$ increases, the energy of this state, shown by the black curve in Figure \ref{fig:energyN2step}(a), decreases, reaches a minimum at $\beta\approx 2.73$ and then starts to slowly grow. When $\beta$ becomes larger than $2.5559$, the role of the ground state $({\cal L},s)=(1,1)$ is transferred to the state $({\cal L},s)=(3,1)$, shown by the green solid curve in Figure \ref{fig:energyN2step}(a). The energies of the states with ${\cal L}=2$ and ${\cal L}=4$ are higher, at all $\beta$, than the energies of the states with ${\cal L}=1$ and ${\cal L}=3$, but the states $({\cal L},s)=(2,1)$ (the red dashed curve) and $({\cal L},s)=(4,1)$ (the blue dashed curve) may be the first excited state in certain intervals of the magnetic field, see Table \ref{tab:N2LtotGS1st}. For all ground and first excited states the value of the second quantum number is $s=1$, i.e. the states with $s>1$ ($s=2$ in the case of two particles) cannot be the first excited state at any $\beta$. This feature remains valid for any $N$. Figure \ref{fig:energyN2step}(b) shows the energy gap between the first excited and the ground state as a function of the magnetic field parameter $\beta$. The gap vanishes at $\beta=2.5559$ and is about $(0.1-0.2)e^2/a_0$ in the range of $1\le \beta\le 4$. 

\begin{table}[!ht]
\caption{The total angular momenta of the ground state ${\cal L}_{\rm GS}$ and of the first excited state ${\cal L}_{\rm 1st}$ assume the values shown in the last two columns in the intervals from $\beta_{\rm from}$ to $\beta_{\rm to}$ shown in the first two columns. The number of particles is  $N=2$, the density profile is step-like. \label{tab:N2LtotGS1st}}
\begin{tabular}{cccc}
$\beta_{\rm from}$ & $\beta_{\rm to}$   & ${\cal L}_{\rm GS}$   & ${\cal L}_{\rm 1st}$\\
\hline
1.0000	& 1.5986	& 1	& 2 \\
1.5986	& 2.5559	& 1	& 3 \\
2.5559	& 3.3567	& 3	& 1 \\
3.3567	& 4.0000	& 3	& 4 \\
\end{tabular}
\end{table}

The energy of the $1/3$ Laughlin state is shown in Figure \ref{fig:energyN2step}(a) by a small red circle at $\beta=3$. At $N=2$ the energy difference between the Laughlin ($-1.871568217$) and the true ground state ($-1.872042007$) is rather small ($=0.00047379$, in units $e^2/a_0$). The dimensionless $\eta$-parameter
\be 
\eta=\frac{E_{\rm RL}-E_{\rm GS}}{E_{\rm 1st}-E_{\rm GS}}\label{eta}
\ee
which characterizes the deviation of the Laughlin energy from the ground state in units of the energy gap at $\beta=3$ equals $\eta=0.004785844574$ at $N=2$.

\subsubsection{Three particles}

\begin{figure}[ht!]
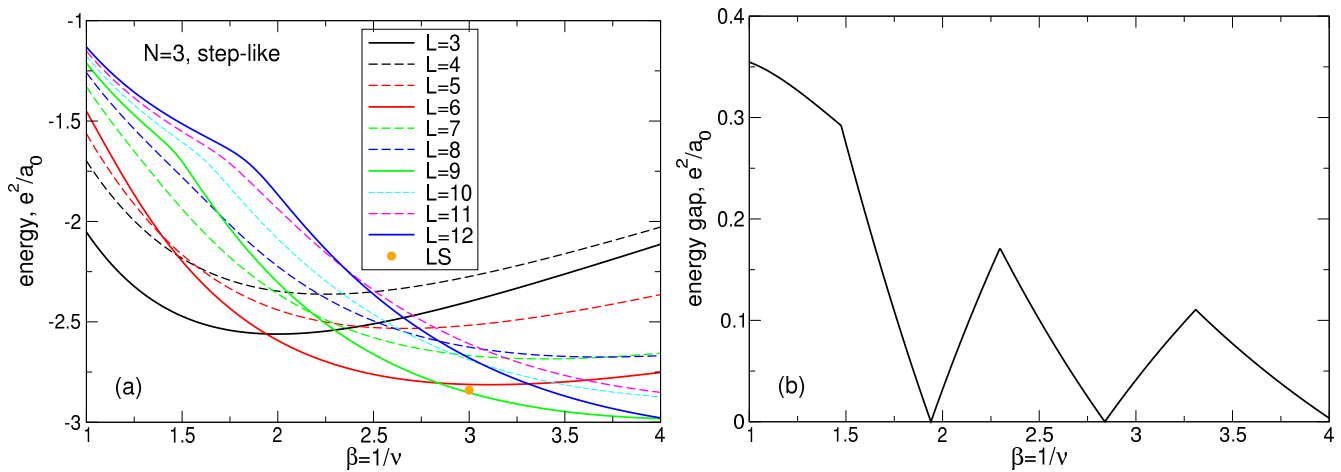

\includegraphics[width=0.49\columnwidth]{N3step.eps}
\includegraphics[width=0.49\columnwidth]{N3stepGap.eps}
\caption{\label{fig:energyN3step} (a) The energy of the many-body states with the total angular momenta from ${\cal L}={\cal L}_{\min}=3$ up to ${\cal L}={\cal L}_{\max}=12$ in the system of $N=3$ 2D electrons as a function of the magnetic field parameter $\beta=1/\nu$. For all shown states the index $s$ equals $s=1$; the states with $s>1$ are not shown. The LS energy at $\nu=1/3$ is shown by a small orange circle. (b) The energy gap between the ground and the first excited states as a function of $\beta$. The positive background density profile is step-like.}
\end{figure}

If $N=3$, the relevant total angular momenta lie between ${\cal L}_{\min}=3$ and ${\cal L}_{\max}=12$, see Table \ref{tab:N3configs}. Figure \ref{fig:energyN3step}(a) shows the energies $E_{{\cal L},s}(\beta)$ for all many-body states with ${\cal L}$ varying from 3 to 12 and for $s=1$. The states with $s\ge 2$ are not shown, since they are neither the ground state nor the first excited state for any $\beta$. The behavior of the curves $E_{{\cal L},1}(\beta)$ is similar to that of the curves for $N=2$. The MDD configuration $|0,1,2\rangle $ with $({\cal L},s)=(3,1)$ remains the ground state from $\beta=1$ until $\beta=1.9397$ and then transfers its role of the ground state to the state $({\cal L},s)=(6,1)$. After $\beta=2.8392$ the state $({\cal L},s)=(9,1)$ become the ground state, see Table \ref{tab:N3LtotGS1st}. The angular momenta values corresponding to the ground states in different $\beta$ intervals (${\cal L}=3$, 6, 9) satisfy the rule
\be 
{\cal L}_j^{\rm GS}={\cal L}_{\min}+Nj, \label{GSrule}
\ee
with integer $j=0,1,2,\dots$. The rule (\ref{GSrule}) is also valid for $N=2$. 

\begin{table}[ht!]
\caption{Possible many-body configurations in a system of $N=3$ particles. $N_{mbs}(N,{\cal L})$ is the total number of all many-particle configurations with given $N$ and ${\cal L}$. \label{tab:N3configs}}
\begin{tabular}{clc}
${\cal L}$ & \hspace{5.5cm} Configurations  & $N_{mbs}$\\
\hline
3 & $|0,1,2\rangle $ & 1\\
4 & $|0,1,3\rangle $ & 1\\
5 & $|0,1,4\rangle $ $|0,2,3\rangle $ & 2 \\
6 & $|0,1,5\rangle $ $|0,2,4\rangle $ $|1,2,3\rangle $ & 3 \\
7 & $|0,1,6\rangle $ $|0,2,5\rangle $ $|0,3,4\rangle $ $|1,2,4\rangle $ & 4 \\
8 & $|0,1,7\rangle $ $|0,2,6\rangle $ $|0,3,5\rangle $ $|1,2,5\rangle $ $|1,3,4\rangle $ & 5 \\
9 & $|0,1,8\rangle $ $|0,2,7\rangle $ $|0,3,6\rangle $ $|0,4,5\rangle $ $|1,2,6\rangle $ $|1,3,5\rangle $ $|2,3,4\rangle $ & 7 \\
10 & $|0,1,9\rangle $ $|0,2,8\rangle $ $|0,3,7\rangle $ $|0,4,6\rangle $ $|1,2,7\rangle $ $|1,3,6\rangle $ $|1,4,5\rangle $ $|2,3,5\rangle $ & 8 \\
11 & $|0,1,10\rangle $ $|0,2,9\rangle $ $|0,3,8\rangle $ $|0,4,7\rangle $  $|0,5,6\rangle $ $|1,2,8\rangle $ $|1,3,7\rangle $ $|1,4,6\rangle $ $|2,3,6\rangle $ $|2,4,5\rangle $ & 10 \\
12 & $|0,1,11\rangle $ $|0,2,10\rangle $ $|0,3,9\rangle $ $|0,4,8\rangle $  $|0,5,7\rangle $ $|1,2,9\rangle $ $|1,3,8\rangle $ $|1,4,7\rangle $ $|1,5,6\rangle $ $|2,3,7\rangle $ $|2,4,6\rangle $ $|3,4,5\rangle $ & 12 \\
\end{tabular}
\end{table}

\begin{table}[!ht]
\caption{The total angular momenta of the ground state ${\cal L}_{\rm GS}$ and of the first excited state ${\cal L}_{\rm 1st}$ assume the values shown in the last two columns in the intervals from $\beta_{\rm from}$ to $\beta_{\rm to}$ shown in the first two columns. The number of particles is  $N=3$, the density profile is step-like. \label{tab:N3LtotGS1st}}
\begin{tabular}{cccc}
$\beta_{\rm from}$ & $\beta_{\rm to}$   & ${\cal L}_{\rm GS}$   & ${\cal L}_{\rm 1st}$\\
\hline
1.0000	& 1.4751	& 3	& 4 \\
1.4751	& 1.9397	& 3	& 6 \\
1.9397	& 2.2971	& 6	& 3 \\
2.2971	& 2.8392	& 6	& 9 \\
2.8392	& 3.3096	& 9	& 6 \\
3.3096	& 4.0000	& 9	& 12 \\
\end{tabular}
\end{table}

Figure \ref{fig:energyN3step}(b) shows the energy gap between the first excited and the ground state as a function of the magnetic field parameter $\beta$. The gap vanishes at two $\beta$-points and is about $(0.05-0.15)e^2/a_0$ in the range of $1\le \beta\le 4$. 

The energy of the $1/3$ Laughlin state is shown by a small orange circle at $\beta=3$. For $N=3$ the deviation of the Laughlin energy ($-2.840219085$) from the true ground state energy ($-2.852910902$) is better seen in the Figure \ref{fig:energyN3step}(a). This difference equals $E_{\rm RL}-E_{\rm GS}=0.012691817250$ in units $e^2/a_0$. The $\eta$-parameter (\ref{eta}) equals $\eta=0.308135289703$ at $N=3$.

\subsubsection{Four particles}

\begin{figure}[ht!]
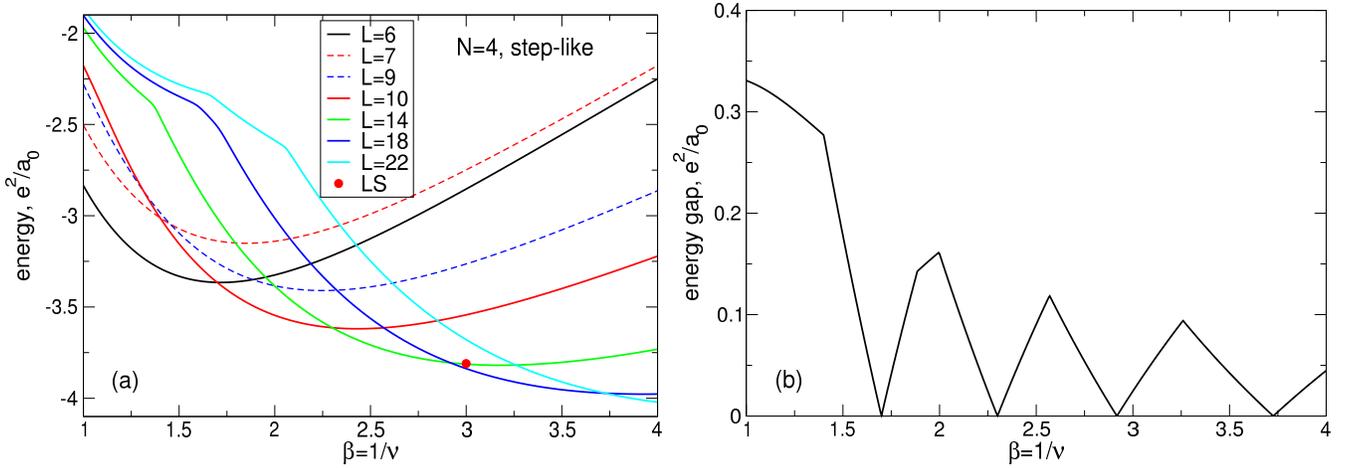

\includegraphics[width=0.49\columnwidth]{N4step.eps}
\includegraphics[width=0.49\columnwidth]{N4stepGap.eps}
\caption{\label{fig:energyN4step} (a) The energy of the many-body states with the total angular momenta from ${\cal L}={\cal L}_{\min}=6$ up to ${\cal L}=22$ in the system of $N=4$ 2D electrons as a function of the magnetic field parameter $\beta=1/\nu$. For all shown states the index $s$ equals $s=1$. Only the states which are either ground or first excited states are shown. The LS energy at $\nu=1/3$ is shown by a small red circle. (b) The energy gap between the ground and the first excited states as a function of $\beta$. The positive background density profile is step-like.}
\end{figure}

If $N=4$, the minimal and maximal total angular momenta (\ref{MinMaxMomenta}) are ${\cal L}_{\min}=6$ and ${\cal L}_{\max}=24$. In order not to overload the graph with too many curves, I show in Figure \ref{fig:energyN4step}(a) only the energies of the states $({\cal L},s)$ that are either the ground or the first excited state in any interval of $\beta$. The rule (\ref{GSrule}) for the ground-states angular momenta remains valid for $N=4$. The $({\cal L},s)$-states with ${\cal L}=8$, $11- 13$, $15-17$, $19-21$, $23-24$, and $s=1$, as well as all states with $s>1$, which can be only the second or higher excited state, are not shown in the Figure. The angular momenta of the ground and first excited states, with the corresponding $\beta$ intervals, are shown in Table \ref{tab:N4LtotGS1st}. Figure \ref{fig:energyN4step}(b) shows the energy gap between the first excited and the ground state as a function of the magnetic field parameter $\beta$. 

\begin{table}[!ht]
\caption{The total angular momenta of the ground state ${\cal L}_{\rm GS}$ and of the first excited state ${\cal L}_{\rm 1st}$ assume the values shown in the last two columns in the intervals from $\beta_{\rm from}$ to $\beta_{\rm to}$ shown in the first two columns. The number of particles is  $N=4$, the density profile is step-like. \label{tab:N4LtotGS1st}}
\begin{tabular}{cccc}
$\beta_{\rm from}$ & $\beta_{\rm to}$   & ${\cal L}_{\rm GS}$   & ${\cal L}_{\rm 1st}$\\
\hline
1.0000	& 1.4006	& 6	& 7 \\
1.4006	& 1.6991	& 6	& 10 \\
1.6991	& 1.8850	& 10	& 6 \\
1.8850	& 1.9972	& 10	& 9 \\
1.9972	& 2.2994	& 10	& 14 \\
2.2994	& 2.5689	& 14	& 10 \\
2.5689	& 2.9173	& 14	& 18 \\
2.9173	& 3.2593	& 18	& 14 \\
3.2593	& 3.7263	& 18	& 22 \\
3.7263	& 4.0000	& 22	& 18 \\
\end{tabular}
\end{table}

The energy of the $1/3$ Laughlin state is shown by a small red circle at $\beta=3$ in Figure \ref{fig:energyN4step}(a). One sees that now, for $N=4$, the energy of the Laughlin state ($-3.809984254$) is not only larger than the energy of the ground state  ($-3.837654843$) but also larger than the energy of the first excited state  ($-3.813057713$). The $\eta$-parameter (\ref{eta}) is larger than one at $N=4$ and equals $\eta=1.124951928311$.

\subsubsection{Five particles}

\begin{figure}[ht!]
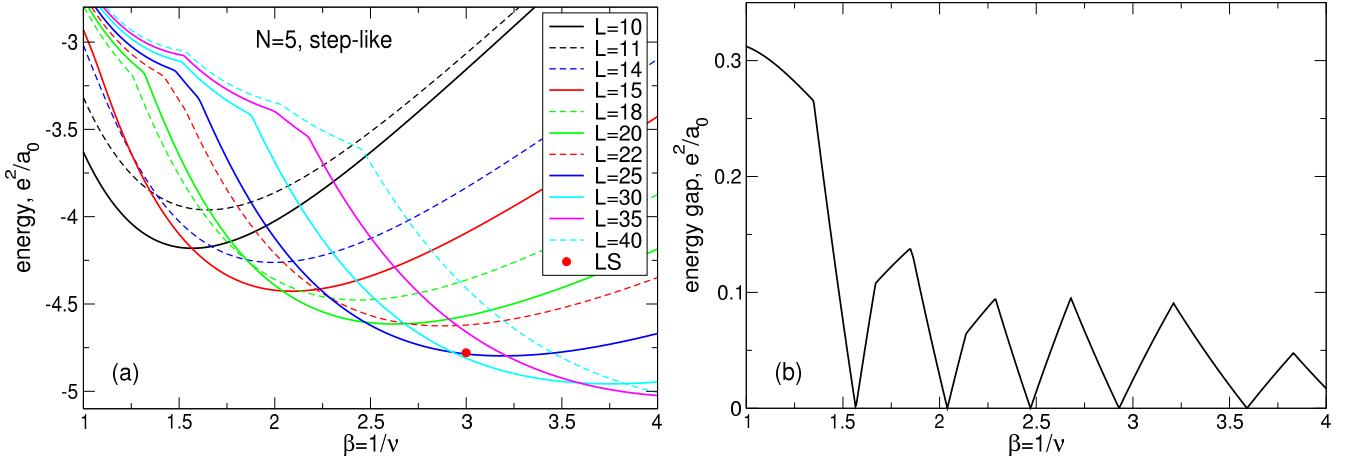

\includegraphics[width=0.49\columnwidth]{N5step.eps}
\includegraphics[width=0.49\columnwidth]{N5stepGap.eps}
\caption{\label{fig:energyN5step} (a) The energy of the many-body states with the total angular momenta from ${\cal L}={\cal L}_{\min}=10$ up to ${\cal L}={\cal L}_{\max}=40$ in the system of $N=5$ 2D electrons as a function of the magnetic field parameter $\beta=1/\nu$. For all shown states the index $s$ equals $s=1$. Only the states which are either ground or first excited states are shown. The LS energy at $\nu=1/3$ is shown by a small red circle. (b) The energy gap between the ground and the first excited states as a function of $\beta$. The positive background density profile is step-like.}
\end{figure}

As in the case of $N=4$, Figure \ref{fig:energyN5step}(a) shows the energies $E_{{\cal L},1}(\beta)$ only for those ${\cal L}$ that are either the ground state or the first excited state in some intervals of $\beta$. These intervals and the corresponding angular momenta ${\cal L}_{\rm GS}$ and ${\cal L}_{\rm 1st}$ are shown in Table \ref{tab:N5LtotGS1st}. The rule (\ref{GSrule}) for ${\cal L}_j^{\rm GS}$ remains valid for $N=5$. Figure \ref{fig:energyN5step}(b) shows the energy gap between the first excited and the ground state as a function of the magnetic field parameter $\beta$. 

\begin{table}[!ht]
\caption{The total angular momenta of the ground state ${\cal L}_{\rm GS}$ and of the first excited state ${\cal L}_{\rm 1st}$ assume the values shown in the last two columns in the intervals from $\beta_{\rm from}$ to $\beta_{\rm to}$ shown in the first two columns. The number of particles is  $N=5$, the density profile is step-like. \label{tab:N5LtotGS1st}}
\begin{tabular}{cccc}
$\beta_{\rm from}$ & $\beta_{\rm to}$   & ${\cal L}_{\rm GS}$   & ${\cal L}_{\rm 1st}$\\
\hline
1.0000	& 1.3483	& 10	& 11 \\
1.3483	& 1.5666	& 10	& 15 \\
1.5666	& 1.6688	& 15	& 10 \\
1.6688	& 1.8486	& 15	& 14 \\
1.8486	& 1.8647	& 15	& 18 \\
1.8647	& 2.0402	& 15	& 20 \\
2.0402	& 2.1372	& 20	& 15 \\
2.1372	& 2.2887	& 20	& 18 \\
2.2887	& 2.2890	& 20	& 22 \\
2.2890	& 2.4705	& 20	& 25 \\
2.4705	& 2.6791	& 25	& 20 \\
2.6791	& 2.9287	& 25	& 30 \\
2.9287	& 3.2100	& 30	& 25 \\
3.2100	& 3.5904	& 30	& 35 \\
3.5904	& 3.8305	& 35	& 30 \\
3.8305	& 4.0000	& 35	& 40 \\
\end{tabular}
\end{table}

The energy of the $1/3$ Laughlin state is shown by a small red circle at $\beta=3$ in Figure \ref{fig:energyN5step}(a). One sees that, as in the previous case, the energy of the Laughlin state ($-4.779626972$) at $N=5$ is also larger than the energy of the first excited state  ($-4.786287998$). The energy of the ground state at $N=5$ is $-4.810805542$, and the $\eta$-parameter (\ref{eta}) equals $\eta=1.271684060185$.

\subsubsection{Six particles}

\begin{figure}[ht!]
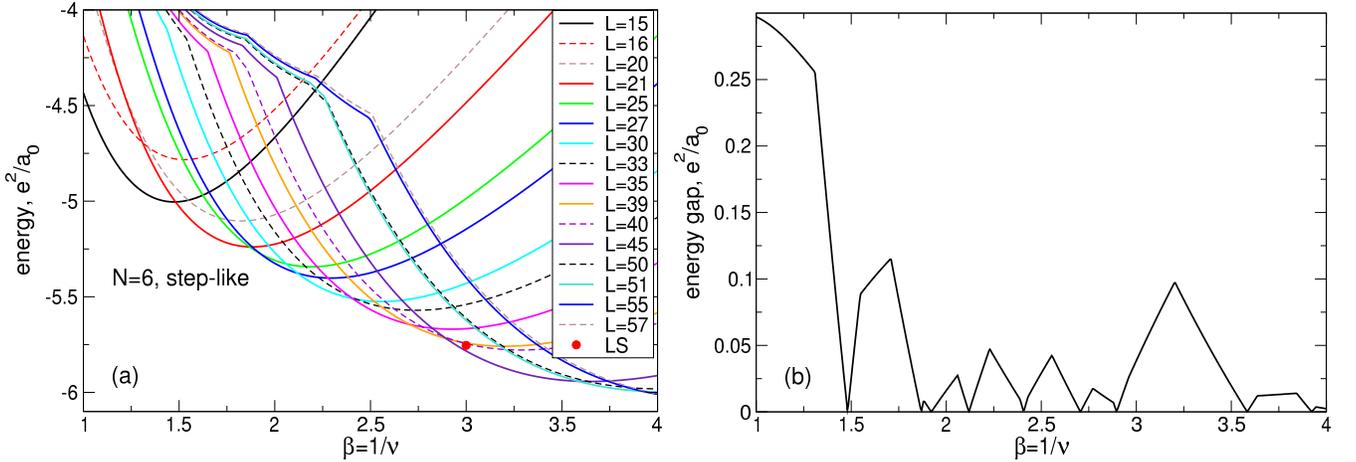

\includegraphics[width=0.49\columnwidth]{N6step.eps}
\includegraphics[width=0.49\columnwidth]{N6stepGap.eps}
\caption{\label{fig:energyN6step} (a) The energy of the many-body states with the total angular momenta from ${\cal L}={\cal L}_{\min}=15$ up to ${\cal L}=57$ in the system of $N=6$ 2D electrons as a function of the magnetic field parameter $\beta=1/\nu$. For all shown states the index $s$ equals $s=1$. Only the states which are either ground or first excited states are shown. The LS energy at $\nu=1/3$ is shown by a small red circle. (b) The energy gap between the ground and the first excited states as a function of $\beta$. The positive background density profile is step-like.}
\end{figure}

For $N=6$, Figure \ref{fig:energyN6step}(a) shows the energies $E_{{\cal L},1}(\beta)$ for those ${\cal L}$ which are either the ground state or the first excited state in some intervals of $\beta$. These intervals and the corresponding angular momenta ${\cal L}_{\rm GS}$ and ${\cal L}_{\rm 1st}$ are shown in Table \ref{tab:N6LtotGS1st}. Figure \ref{fig:energyN6step}(b) shows the energy gap between the first excited and the ground state as a function of the magnetic field parameter $\beta$. 

\begin{table}[!ht]
\caption{The total angular momenta of the ground state ${\cal L}_{\rm GS}$ and of the first excited state ${\cal L}_{\rm 1st}$ assume the values shown in the last two columns in the intervals from $\beta_{\rm from}$ to $\beta_{\rm to}$ shown in the first two columns. The number of particles is  $N=6$, the density profile is step-like. \label{tab:N6LtotGS1st}}
\begin{tabular}{cccc}
$\beta_{\rm from}$ & $\beta_{\rm to}$   & ${\cal L}_{\rm GS}$   & ${\cal L}_{\rm 1st}$\\
\hline
1.0000	& 1.3094	& 15	& 16 \\
1.3094	& 1.4812	& 15	& 21 \\
1.4812	& 1.5480	& 21	& 15 \\
1.5480	& 1.7089	& 21	& 20 \\
1.7089	& 1.8685	& 21	& 25 \\
1.8685	& 1.8813	& 25	& 21 \\
1.8813	& 1.9215	& 25	& 27 \\
1.9215	& 2.0598	& 27	& 25 \\
2.0598	& 2.1190	& 27	& 30 \\
2.1190	& 2.2284	& 30	& 27 \\
2.2284	& 2.3883	& 30	& 33 \\
2.3883	& 2.4072	& 30	& 35 \\
2.4072	& 2.4303	& 35	& 30 \\
2.4303	& 2.5560	& 35	& 33 \\
2.5560	& 2.7060	& 35	& 39 \\
2.7060	& 2.7714	& 39	& 35 \\
2.7714	& 2.8798	& 39	& 40 \\
2.8798	& 2.8965	& 39	& 45 \\
2.8965	& 2.9606	& 45	& 39 \\
2.9606	& 3.2029	& 45	& 40 \\
3.2029	& 3.5837	& 45	& 51 \\
3.5837	& 3.6374	& 51	& 45 \\
3.6374	& 3.8437	& 51	& 50 \\
3.8437	& 3.9236	& 51	& 55 \\
3.9236	& 3.9456	& 55	& 51 \\
3.9456	& 4.0000	& 55	& 57 \\
\end{tabular}
\end{table}

In the case of six particles, the overall picture of the energy spectra turns out to be more complex than for smaller $N$. Firstly, the rule (\ref{GSrule}) is no longer satisfied for $N=6$. Secondly, for $N=3$, 4 and 5, the number of ${\cal L}$-states, which were the ground state in the interval $1\le\beta\le 4$, was equal to 4, 5 and 6, i.e. $N+1$, so that one would expect that at ${\cal L}=6$ this number would be 7. But calculations show that at $N=6$ ten different ${\cal L}$-states can take on the role of the ground state. The reasons of these two features of the six-particle system are discussed below in Section \ref{sec:prelim-sum}.

Another noticeable result of the $N=6$ case is that the largest energy gap and the largest distance between the gap nodes are seen in Figure \ref{fig:energyN6step}(b) around $\beta=1.5$ and $\beta=3$ which correspond to $\nu=2/3$ and $\nu=1/3$. So, some correlations with the FQHE experiment arise as $N$ increases. 

The energy of the $1/3$ Laughlin state is shown by a small red circle at $\beta=3$ in Figure \ref{fig:energyN6step}(a). In the case of $N=6$ this energy ($-5.754096859$) is larger than the energy of the ground state  ($-5.783596891$) but smaller than the energy of the first excited state  ($-5.745154444$). The $\eta$-parameter (\ref{eta}) for $N=6$ equals $\eta=0.767381745006$.

\subsubsection{Seven particles, step-like density profile}

\begin{figure}[ht!]
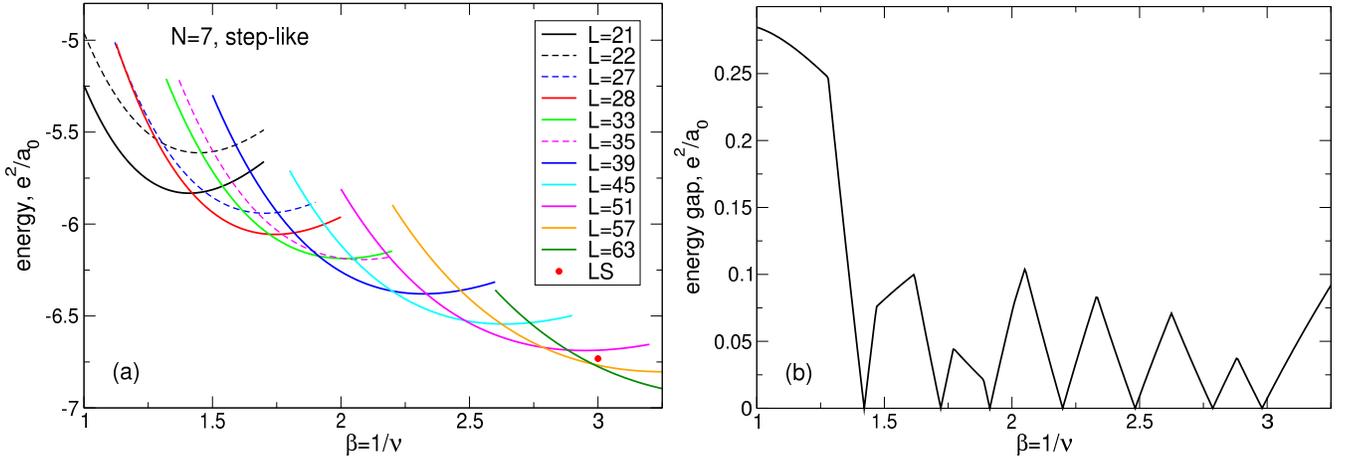

\includegraphics[width=0.49\columnwidth]{N7step.eps}
\includegraphics[width=0.49\columnwidth]{N7stepGap.eps}
\caption{\label{fig:energyN7step} (a) The energy of the many-body states with the total angular momenta from ${\cal L}={\cal L}_{\min}=21$ up to ${\cal L}=63$ in the system of $N=7$ 2D electrons as a function of the magnetic field parameter $\beta=1/\nu$. For all shown states the index $s$ equals $s=1$. Only the states which are either ground or first excited states are shown. The LS energy at $\nu=1/3$ is shown by a small red circle. (b) The energy gap between the ground and the first excited states as a function of $\beta$. The positive background density profile is step-like.}
\end{figure}

For $N=7$ the same data as in the previous pictures are shown in Figure \ref{fig:energyN7step} and in Table \ref{tab:N7LtotGS1st} in the interval $1\le\beta\le 3.25$. As compared to the case of six particles the spectra look simpler again, with $N+1=8$ angular momenta, corresponding to the ground state in different intervals of $\beta$. The equidistance rule (\ref{GSrule}) is satisfied with some corrections. Firstly, instead of $Nj$ on the right-hand side of (\ref{GSrule}), the rule for $N=7$ changes to 
\be 
{\cal L}_j^{\rm GS}(N=7)={\cal L}_{\min}+(N-1)j.\label{GSrule7}
\ee
Secondly, the rule (\ref{GSrule7}) would give the following sequence of the ground-state ${\cal L}$-states: \textbf{21}, 27, \textbf{33}, \textbf{39}, \textbf{45}, \textbf{51}, \textbf{57}, and \textbf{63}. The numbers printed in bold here are the angular momenta actually corresponding to the ground state at certain $\beta$ intervals. But instead of ${\cal L}=27$, the ground state in the interval $1.4211\le\beta\le 1.721$ is the state with ${\cal L}=28$. The reasons for these deviations are discussed in the sections \ref{sec:prelim-sum} and \ref{sec:density7step}.

\begin{table}[!ht]
\caption{The total angular momenta of the ground state ${\cal L}_{\rm GS}$ and of the first excited state ${\cal L}_{\rm 1st}$ assume the values shown in the last two columns in the intervals from $\beta_{\rm from}$ to $\beta_{\rm to}$ shown in the first two columns. The number of particles is  $N=7$, the density profile is step-like. \label{tab:N7LtotGS1st}}
\begin{tabular}{cccc}
$\beta_{\rm from}$ & $\beta_{\rm to}$   & ${\cal L}_{\rm GS}$   & ${\cal L}_{\rm 1st}$\\
\hline
1.0000	& 1.2795	& 21	& 22 \\
1.2795	& 1.4211	& 21	& 28 \\
1.4211	& 1.4690	& 28	& 21 \\
1.4690	& 1.6170	& 28	& 27 \\
1.6170	& 1.7210	& 28	& 33 \\
1.7210	& 1.7708	& 33	& 28 \\
1.7708	& 1.8886	& 33	& 35 \\
1.8886	& 1.9132	& 33	& 39 \\
1.9132	& 2.0082	& 39	& 33 \\
2.0082	& 2.0512	& 39	& 35 \\
2.0512	& 2.1988	& 39	& 45 \\
2.1988	& 2.3323	& 45	& 39 \\
2.3323	& 2.4828	& 45	& 51 \\
2.4828	& 2.6249	& 51	& 45 \\
2.6249	& 2.7866	& 51	& 57 \\
2.7866	& 2.8819	& 57	& 51 \\
2.8819	& 2.9803	& 57	& 63 \\
2.9803	& 3.2500	& 63	& 57 \\
\end{tabular}
\end{table}

The energy of the $1/3$ Laughlin state is shown by a small red circle at $\beta=3$ in Figure \ref{fig:energyN7step}(a). One sees that in the case of $N=7$ particles the ground (${\cal L}=63$) and the first excited states (${\cal L}=57$) are very close in energy, $E_{\rm GS}=-6.775064708$ and $E_{\rm 1st}=
-6.7677608$, respectively, while the energy of the Laughlin state, $E_{\rm RL}=-6.732099378$ is substantially larger than both $E_{\rm GS}$ and $E_{\rm 1st}$. As a result, the $\eta$-parameter (\ref{eta}) in the case of $N=7$ is much larger than one, $\eta=5.882512643180$.

\subsubsection{Seven particles, smooth density profile\label{sec:resultsSmooth}}

\begin{figure}[ht!]
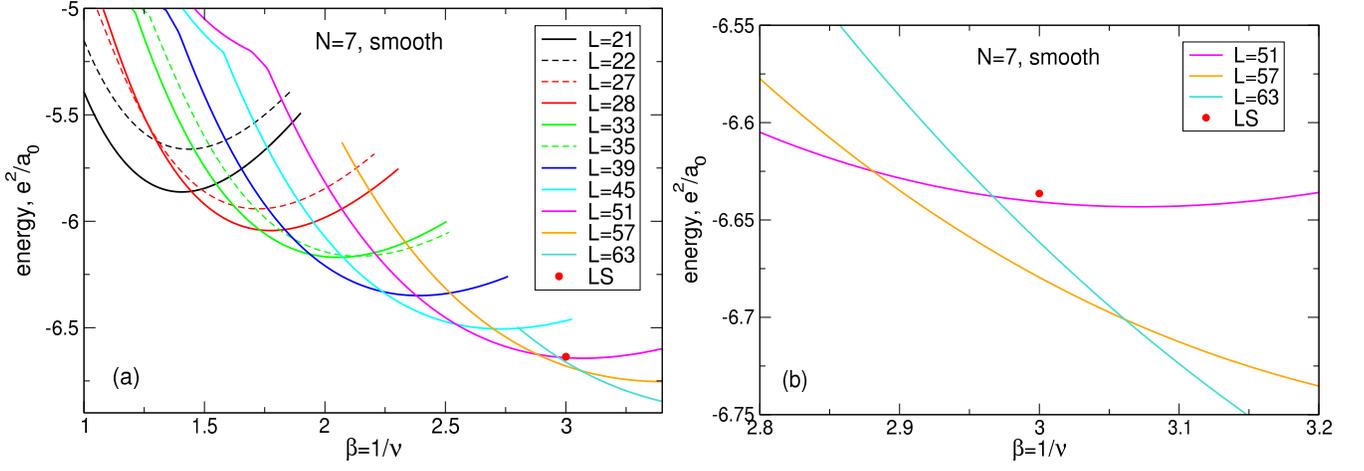

\includegraphics[width=0.49\columnwidth]{N7smoo.eps}
\includegraphics[width=0.49\columnwidth]{N7smooAtb=3.eps}
\caption{\label{fig:energyN7smoo} (a) The energy of the many-body states with the total angular momenta from ${\cal L}={\cal L}_{\min}=21$ up to ${\cal L}=63$ in the system of $N=7$ 2D electrons as a function of the magnetic field parameter $\beta=1/\nu$. For all shown states the index $s$ equals $s=1$. Only the states which are either ground or first excited states are shown. The Laughlin state energy at $\nu=1/3$ is shown by a small red circle. (b) The vicinity of the point $\beta=3$ on a larger scale. The positive background density profile is smooth.}
\end{figure}

It is interesting to check, whether and how the energy spectra are modified if the density profile is smooth. Figure \ref{fig:energyN7smoo}(a) shows the energy of the ground and the first excited states as a function of $\beta$ for the case of the smooth density profile (\ref{dens-smooth}). In general, the curves $E_{\rm GS}(\beta)$ and $E_{\rm 1st}(\beta)$ look similar to the case of the step-like profile, but some details are quantitatively different. In particular, the intervals of $\beta$ corresponding to different ${\cal L}$ values differ from the case of the step-like profile, see Table \ref{tab:N7LtotGS1stSmooth}. 

\begin{table}[!ht]
\caption{The total angular momenta of the ground state ${\cal L}_{\rm GS}$ and of the first excited state ${\cal L}_{\rm 1st}$ assume the values shown in the last two columns in the intervals from $\beta_{\rm from}$ to $\beta_{\rm to}$ shown in the first two columns. The number of particles is  $N=7$, the density profile is smooth. \label{tab:N7LtotGS1stSmooth}}
\begin{tabular}{cccc}
$\beta_{\rm from}$ & $\beta_{\rm to}$   & ${\cal L}_{\rm GS}$   & ${\cal L}_{\rm 1st}$\\
\hline
   1.0000 &    1.3016 &   21 &   22 \\
   1.3016 &    1.4455 &   21 &   28 \\
   1.4455 &    1.4849 &   28 &   21 \\
   1.4849 &    1.6180 &   28 &   27 \\
   1.6180 &    1.7293 &   28 &   33 \\
   1.7293 &    1.8142 &   33 &   28 \\
   1.8142 &    1.8689 &   33 &   35 \\
   1.8689 &    1.9380 &   33 &   39 \\
   1.9380 &    2.0744 &   39 &   33 \\
   2.0744 &    2.2388 &   39 &   45 \\
   2.2388 &    2.3790 &   45 &   39 \\
   2.3790 &    2.5443 &   45 &   51 \\
   2.5443 &    2.6990 &   51 &   45 \\
   2.6990 &    2.8809 &   51 &   57 \\
   2.8809 &    2.9667 &   57 &   51 \\
   2.9667 &    3.0613 &   57 &   63 \\
   3.0613 &    3.4000 &   63 &   57 \\
\end{tabular}
\end{table}

A significant difference from the case of the step-like profile is observed in the energy of the Laughlin state, Figure \ref{fig:energyN7smoo}(b). In the case of the step-like profile, Figure \ref{fig:energyN7step}(a), the ground, first excited and second excited states at $\beta=3$ had the angular momenta ${\cal L}_{\rm GS}=63$, ${\cal L}_{\rm 1st}=57$, and ${\cal L}_{\rm 2nd}=51$, respectively. The energy of the Laughlin state lay between the energies of the first and second excited states. In the case of the smooth density profile, Figure \ref{fig:energyN7smoo}(b), the ground, first and second excited states at $\beta=3$ have the angular momenta ${\cal L}_{\rm GS}=57$, ${\cal L}_{\rm 1st}=63$, and ${\cal L}_{\rm 2nd}=51$, respectively, and the energy of the Laughlin state not only exceeds the energy of the first excited state, but lie above the energy of the second excited state, see Figure \ref{fig:energyN7smoo}(b) and Table \ref{tab:N7EnergyGS1st2ndLauSmooth}. 

\begin{table}[!ht]
\caption{The energies of the ground state, first and second excited states, as well as of the Laughlin state, at $\beta=1/\nu =3$, in the case of the smooth density profile. \label{tab:N7EnergyGS1st2ndLauSmooth}}
\begin{tabular}{l|c|c|c}
\hspace{3mm}State & \hspace{2mm}${\cal L}$ \hspace{2mm}  & \hspace{5mm}$E_{\rm state}$ \hspace{5mm}  & \hspace{5mm}$E_{\rm state}-E_{\rm GS}$\hspace{5mm}\\
\hline
Laughlin & &	-6.6363835 &	0.0431284 \\
2nd excited &	51&	-6.6407409 & 0.0387709 \\
1st excited &	63	& -6.6613834 &	0.0181285 \\
Ground  &	57	& -6.6795119 & 0.0 \\
\end{tabular}
\end{table}

\subsubsection{Preliminary summary and discussion\label{sec:prelim-sum}}

The results for the energy spectra obtained above show that
\begin{enumerate}
\item the ground state energy decreases with increasing magnetic field, oscillating due to a stepwise increase of the total angular momentum ${\cal L}_{\rm GS}$ in the ground states;
\item the total ground-state angular momenta ${\cal L}_{\rm GS}$ show, as a rule, a certain periodicity with the period 
\be 
\delta{\cal L}=\left\{
\begin{array}{ll}
N, &\textrm{ if }N< 6 \\
N-1, &\textrm{ if }N> 6 \\
\end{array}
\right.,
\label{DeltaLrule}
\ee
for example, at $N=4$ the angular momenta corresponding to the ground states in different $\beta$ intervals are ${\cal L}_{\rm GS}=10,$ 15, 20, 25, 30, 35, etc.  
\item some deviations from the rule (\ref{DeltaLrule}) are seen at $N=6$ and at $N=7$ at low magnetic fields ($\beta\lesssim 1.75$).
\end{enumerate}
What are the reasons for the rule (\ref{DeltaLrule}) and deviations from it? 

\begin{figure}[ht!]
\includegraphics[width=0.49\columnwidth]{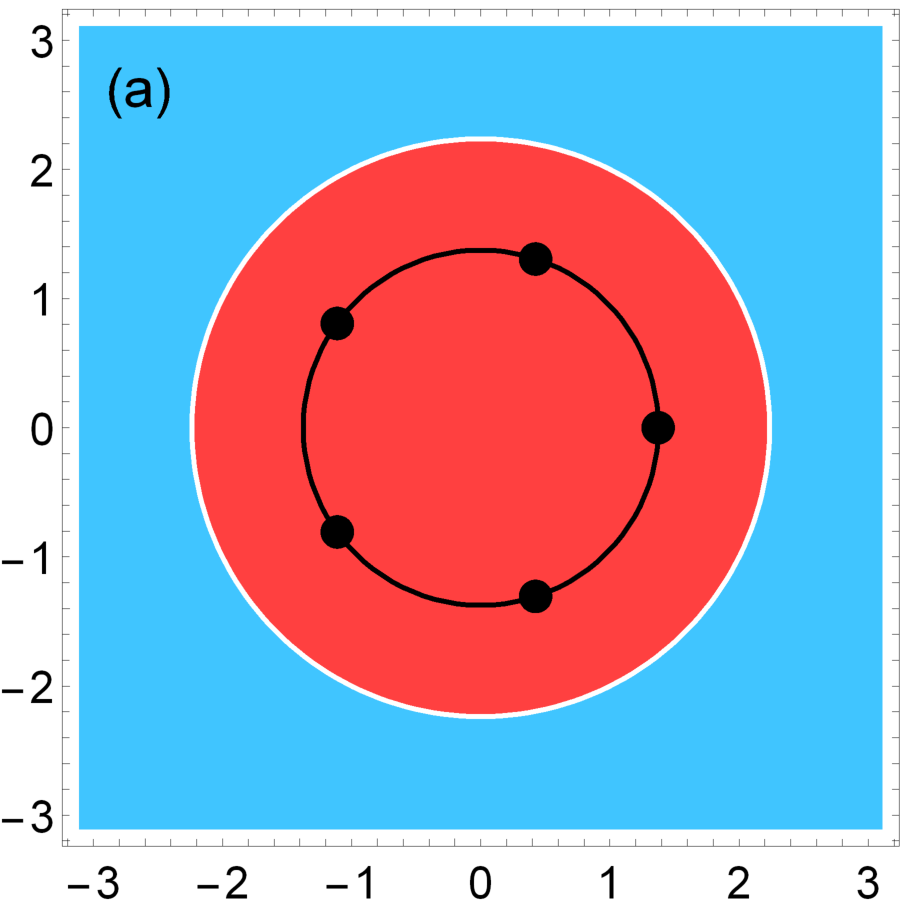}
\includegraphics[width=0.49\columnwidth]{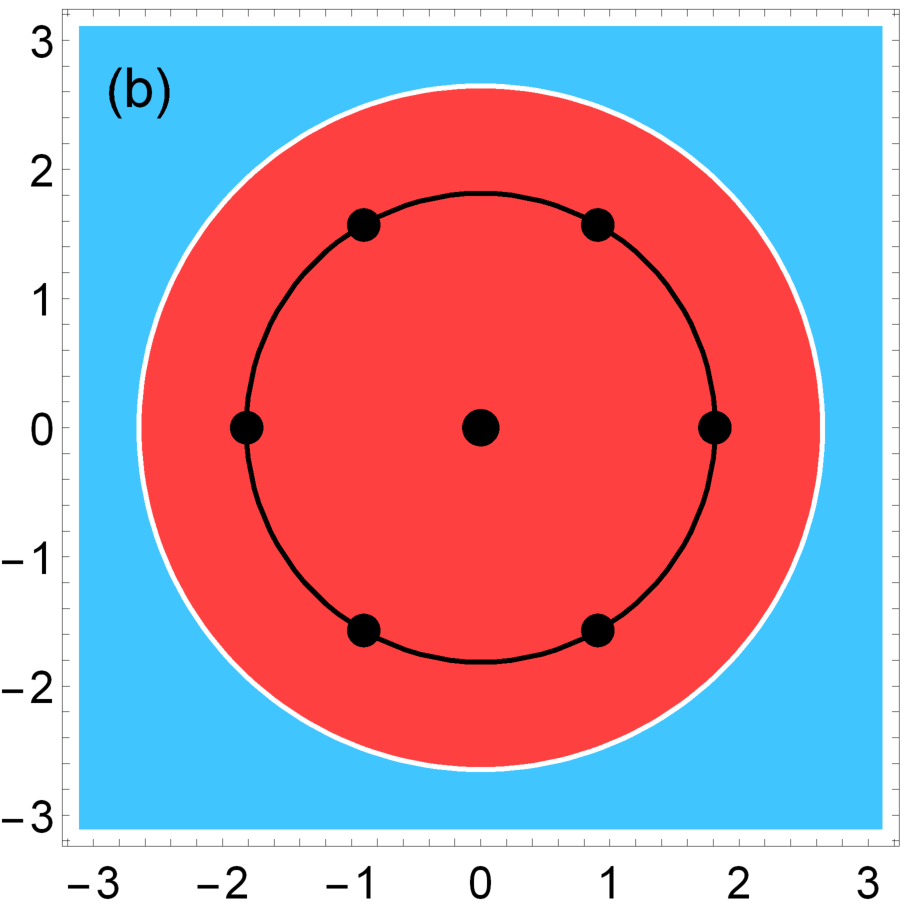}
\caption{\label{fig:WC_N57} The classical Wigner crystal configurations for (a) $N=5$ and (b) $N=7$ particles. The positive background density profile is step-like. The boundary of the positively charged disk is shown by white circles in the pictures.}
\end{figure}

As was discussed in Section III A of Ref. \cite{Mikhailov23a}, at $N\le 8$ classical point charges in the attractive positive-background potential are arranged in two types of spatial configurations: with a single shell $(0,N)$, when all electrons are located on a ring of a finite radius $R_s$, Figure \ref{fig:WC_N57}(a), or with two shells $(1,N-1)$, with one particle at the center of the disk, and $N-1$ particles around the center, Figure \ref{fig:WC_N57}(b). If $N\neq 6$, the number of electrons on the outer shell,
\be 
N_{\rm out-sh}=\left\{ 
\begin{array}{ll}
N, &\textrm{ if }N< 6, \\ 
N-1, & \textrm{ if }N> 6, \\
\end{array}
\right. 
\ee
coincides with $\delta{\cal L}$ in Eq. (\ref{DeltaLrule}). It is therefore reasonable to assume, that due to the periodic arrangement of classical point
particles on the outer shell of the Wigner molecules, the quantum solution, the many-body wave function $|\Psi\rangle$, should have an internal periodicity in polar angle $\phi$ with the period $2\pi/N_{\rm out-sh}$ and, hence, be formed from the states that satisfy the rule (\ref{DeltaLrule}). 

Why is the ${\cal L}_{\rm GS}$ sequence more complicated and deviates significantly from the rule (\ref{DeltaLrule}) in the case of six particles? If $N=6$, the classical energies of both one-shell and two-shells configurations are very close to each other, both in the step-like \cite{Mikhailov23a} and in the smooth density profile models. Therefore, in the quantum-mechanical solution one can expect the ground-state angular momenta both with the period $\delta{\cal L}=N=6$ and with the period $\delta{\cal L}=N-1=5$. In the former case one would have the (``underlined'') sequence
\bc ${\cal L}_{\rm GS}=\underline{15}$, \underline{21}, \underline{27}, \underline{33}, \underline{39}, \underline{45}, \underline{51}, \underline{57}. \ec
In the latter case one would expect 
\bc ${\cal L}_{\rm GS}=15$, 20, 25, 30, 35, 40, 45, 50, 55. \ec
The exact calculation shows that realized is the sequence
\bc ${\cal L}_{\rm GS}=\underline{15}$, \underline{21}, 25, \underline{27}, 30, 35, \underline{39}, \underline{45}, \underline{51}, 55, \ec
while the states with 
\bc ${\cal L}_{\rm 1st}=20$, \underline{33}, 40, 50, \underline{57} \ec
serve as the low-lying first excited states. Thus the complicated structure of levels at $N=6$ is explained by the competition of one-shell and two-shells configurations which have very close energies already in the classical approach.

The second deviation from the rule (\ref{DeltaLrule}) (at $N=7$ and low magnetic fields) can be explained after analysis of the electron densities in the ground and excited states of the system. Such an analysis is performed in the next section.

\subsection{Electron density\label{sec:ResDensity}}

In this Section I analyze the density of electrons $n_e(r)$ in a number of different ${\cal L}$-states for two cases, representing the one-shell and two-shells configurations: $N=5$ and $N=7$.

\subsubsection{Five particles}

The energy spectra of the system of $N=5$ electrons are shown in Figure \ref{fig:energyN5step}(a). One of the states shown there (by the blue solid curve) is characterized by the total angular momentum ${\cal L}=25$. It is the ground state of the five-particle system at $2.4705\le\beta\le 2.9287$, but it is instructive to plot the density of this state in a broader range of $\beta$. 

\begin{figure}[ht!]
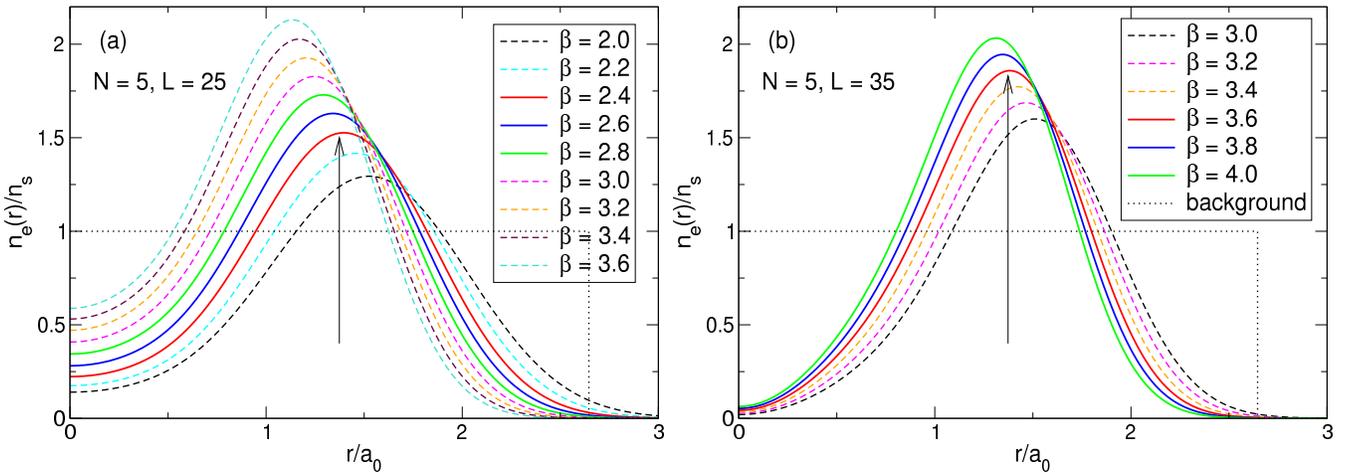

\includegraphics[width=0.49\columnwidth]{DensityN5L25.eps}
\includegraphics[width=0.49\columnwidth]{DensityN5L35.eps}
\caption{\label{fig:Density_N5L2535} The density of electrons in the system of $N=5$ particles at different magnetic fields and at (a) ${\cal L}=25$ and (b) ${\cal L}=35$. For $\beta$-values, for which the corresponding ${\cal L}$ state is the ground state, the function $n_e(r)$ is shown by solid curves. The arrow at $r/a_0=R_s/a_0=1.373422$ shows the position of the shell radius in the classical Wigner molecule. The positive background density profile is step-like and shown by the black dotted curve in both panels.}
\end{figure}

Figure \ref{fig:Density_N5L2535}(a) shows the density of the ${\cal L}=25$ state for $\beta$ varying from $\beta=2.0$ up to $\beta=3.6$ with the step 0.2. At $\beta=2.0$ this is the fifth excited state, and at $\beta=3.6$ it is the third excited state. One sees that at all $\beta$ the density of electrons has the shape of a ring. When $\beta=2.0$ the density maximum equals $n_e/n_s\approx 1.294$ and is located at $r/a_0=R_{\max}/a_0\approx 1.525$, which is larger than the classical shell radius $R_s/a_0=1.373422$. This is not surprising because the state with ${\cal L}=25$ is not the ground state at $\beta=2.0$. Then, when the magnetic field increases, the length $\lambda$ in Eq. (\ref{lambda}) decreases, the wave function shrinks, and the density maximum shifts to smaller values of $r/a_0$. For the values of $\beta=2.4$, 2.6, and 2.8, for which the state ${\cal L}=25$ is the ground state, the maxima of the electron density lie at $R_{\max}/a_0\approx 1.395$, 1.34, and 1.29, respectively. These density maxima are close to or slightly below the classical shell radius $R_s/a_0=1.373422$. As the $\beta$-parameter increases further, the ring radius $R_{\max}$ becomes too small compared to the classical shell radius $R_s$, and the state with ${\cal L}=25$ ceases to be the ground state. Instead, the state with a larger ${\cal L}$ takes on the role of the ground state, since the states with larger ${\cal L}$'s have a larger ring radius. 

Figure \ref{fig:Density_N5L2535}(b) confirms this simple physical picture. It shows the density of electrons for the state with ${\cal L}=35$ and for $\beta$ varying from $\beta=3.0$ up to $\beta=4.0$ with the interval of 0.2. Again, at all values of $\beta$ the density has the shape of a ring, but its radius $R_{\max}$ is larger (at the same $\beta$) then in panel (a) since the angular momentum is now bigger. At $\beta=3.0$ the state with ${\cal L}=35$ is the second excited state and it becomes the ground state in the interval $3.5904\le \beta\le 4.0$. If $\beta=3.0$, the density maximum ($R_{\max}/a_0=1.51$) lies well above the classical shell radius $R_s/a_0=1.373422$. At $\beta=3.2$ and 3.4 the density maxima are at $R_{\max}/a_0=1.46$ and 1.42 respectively, which is still larger than $R_s/a_0$. But when $\beta$ lies in the range where the state  ${\cal L}=35$ is the ground state ($\beta=3.6$, 3.8 and 4.0), the ring radii become almost equal and slightly smaller ($R_{\max}/a_0\approx 1.38$, 1.35, and 1.31, respectively), than the classical shell radius $R_s/a_0=1.373422$. 

Thus, while in the classical solution five point particles are at the distance $R_s$ from the disk center, in the quantum solution the probability to find an electron at a point $\bm r$ has the shape of a ring with the radius $R_{\max}$ being very close to the classical shell radius $R_s$. When the magnetic field changes the quantum ring radius $R_{\max}$ weakly oscillated near the classical shell radius $R_s$.

\subsubsection{Seven particles, step-like density profile\label{sec:density7step}}

Now consider a system of $N=7$ particles which illustrates the case of the two-shells configuration. In addition, the Wigner molecule with seven electrons is a small piece of a macroscopic Wigner crystal, therefore it is useful to study this case in detail. I show results for the ground-state density for all $\beta$ from 1.0 to 3.4, with the step 0.1. The results help not only to understand the physics of the ground state of FQHE systems, but also to establish the limits of applicability of the approximations made.

\begin{figure}[ht!]
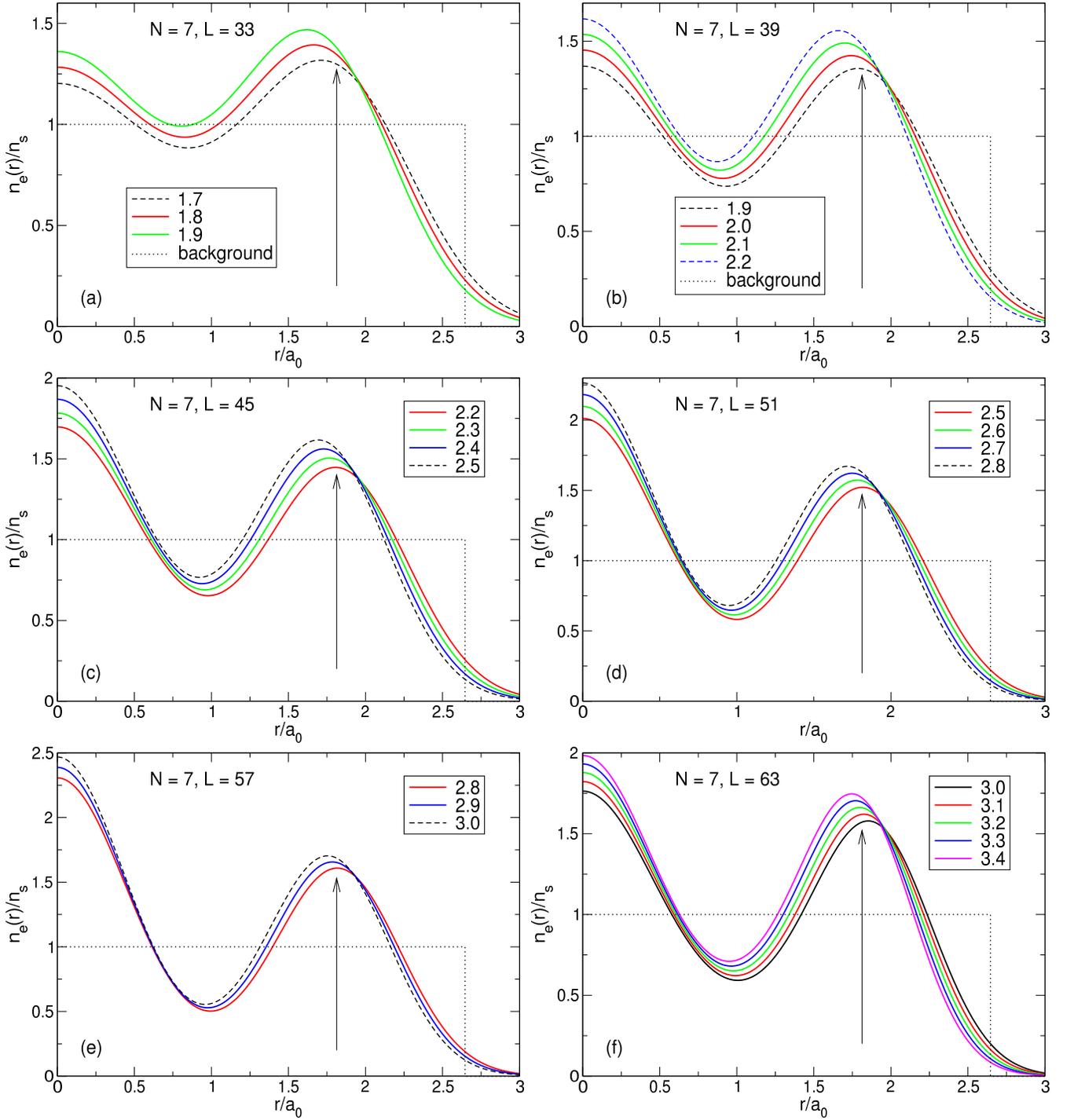

\includegraphics[width=0.49\columnwidth]{DensN7L33.eps}
\includegraphics[width=0.49\columnwidth]{DensN7L39.eps}
\includegraphics[width=0.49\columnwidth]{DensN7L45.eps}
\includegraphics[width=0.49\columnwidth]{DensN7L51.eps}
\includegraphics[width=0.49\columnwidth]{DensN7L57.eps}
\includegraphics[width=0.49\columnwidth]{DensN7L63.eps}
\caption{\label{fig:Density_N7} The density of electrons in the system of $N=7$ particles at different magnetic fields and at (a) ${\cal L}=33$, (b) ${\cal L}=39$, (c) ${\cal L}=45$, (d) ${\cal L}=51$, (e) ${\cal L}=57$, and (f) ${\cal L}=63$. For $\beta$-values, for which the corresponding ${\cal L}$ state is the ground state, the function $n_e(r)$ is shown by solid curves. The arrow at $r/a_0=R_s/a_0=1.8126$ shows the position of the shell radius in the classical Wigner molecule. The step-like positive background density profile is shown by the black dotted curve in all panels.}
\end{figure}

It is instructive to consider first the case of large magnetic fields. Figure \ref{fig:Density_N7} shows the electron density for $\beta\ge 1.7$. The panels (a) to (f) show results for the total angular momenta from ${\cal L}=33$ to ${\cal L}=63$ with the step $\delta{\cal L}=6$. All these states are the ground states in the corresponding magnetic field intervals, see Figure \ref{fig:energyN7step}(a).

\begin{figure}[ht!]
\includegraphics[width=0.49\columnwidth]{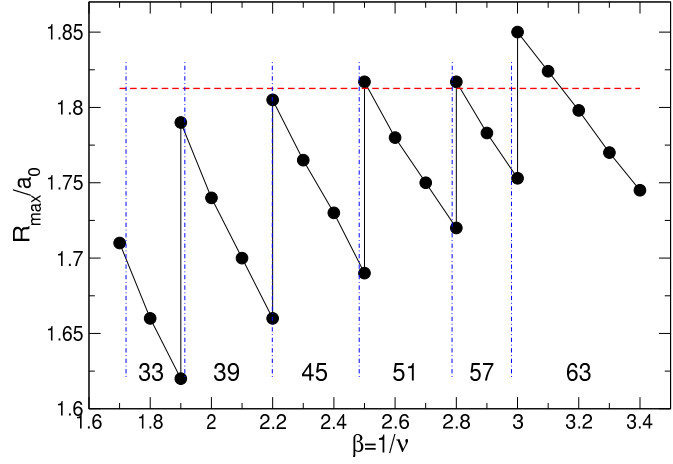}
\caption{\label{fig:DensMaxima} Positions of the maxima of the density curves shown in Figure \ref{fig:Density_N7} at different $\beta$-values and different angular momenta ${\cal L}$. The red dashed line at $R/a_0=1.8126$ shows the position of the shell radius in the classical Wigner molecule. The blue dash-dotted lines separate the areas where the ground states have different angular momenta indicated by the numbers 33, 39, $\dots$, 63 in the corresponding $\beta$-intervals.}
\end{figure}

In full agreement with the classical distribution of point charges, all density curves in Figures \ref{fig:Density_N7}(a)-(f) have two maxima, one at $r=0$ and the other at a finite $r=R_{\max}$. %The finite-$r$ maxima are close to the shell radius of the classical Wigner molecule. 
The width of the maxima is determined by the size $\lambda$ of the wave functions (\ref{spwf_psiL}), $\lambda/a_0=\sqrt{\nu}=1/\sqrt{\beta}$, therefore it  decreases as $\beta$ grows, from panel (a) to panel (f). Due to the same reason, the wave function shrinks with increasing $\beta$, at any given ${\cal L}$, and the positions of maxima of the density curves $R_{\max}$ shift toward smaller $r$-values. The values of $R_{\max}$ are close to or smaller than the classical shell radius $R_s$, see Figure \ref{fig:DensMaxima}. The fact that the quantum-mechanically calculated radii of the rings $R_{\max}$ are, on average, less than the classical shell radii $R_s$ is not surprising, since the potential energy of the attractive background potential is higher at $r>R_s$ and lower at $r<R_s$ compared to $V_b(r=R_s)$, see Figure \ref{fig:Ubpotential}. 

Thus, in strong magnetic fields $\beta>1.7$ the distribution of the electron density in the ground state of the system perfectly agrees with that of the classical Wigner molecule. As magnetic field increases, the outer ring radius oscillates around the classical value $R_s$ gradually decreasing with increasing $B$ and increasing abruptly as the angular momentum changes by $\delta {\cal L}=6$. The number $\delta {\cal L}=6$ is related to the $C_6$ symmetry of the macroscopic Wigner crystal, since the wave function should be periodic in the polar angle $\phi$ with the period $2\pi/\delta {\cal L}$. 

\begin{figure}[ht!]
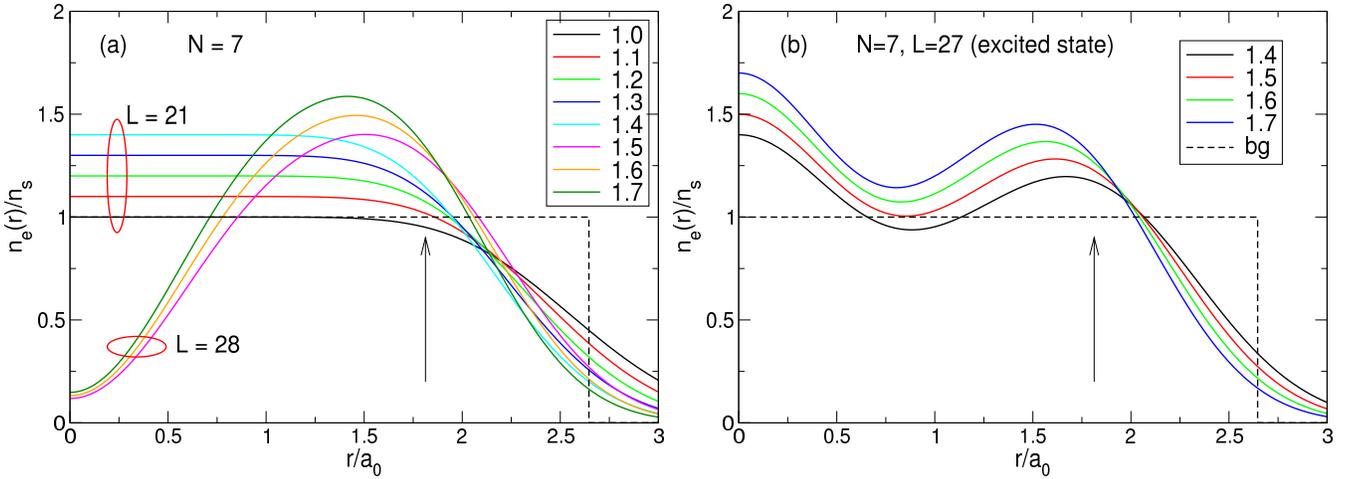

\includegraphics[width=0.49\columnwidth]{DensN7L2128.eps}
\includegraphics[width=0.49\columnwidth]{DensN7L27.eps}
\caption{\label{fig:Density_N7lowB} The density of electrons in the system of $N=7$ particles at different magnetic fields and (a) in the ground states with ${\cal L}=21$ and 28, and (b) in the excited state with ${\cal L}=27$. The arrow at $r/a_0=R_s/a_0=1.8126$ shows the position of the shell radius in the classical Wigner molecule. The step-like positive background density profile is shown by the black dashed curve in all panels.}
\end{figure}

At lower magnetic fields $\beta\lesssim 1.72$, the shape of the electron density in the ground state changes significantly, Figure \ref{fig:Density_N7lowB}(a). Here, one should consider two different situations. If $1.0\le\beta\le 1.4211$, the ground state has the total angular momentum ${\cal L}=21$, the expansion of the many-body wave function (\ref{PsiExpansion}) contains only one Slater determinant, $N_{mbs}=1$, and the wave function is the maximum density droplet state $|0,1,2,3,4,5,6\rangle$. The electron density in this state, Figure \ref{fig:Density_N7lowB}(a), does not have an internal structure typical for a crystal, but has the form of a uniform liquid. The radius of the liquid spot decreases with increasing $\beta$. 

Evidently, this solution is incorrect, since for a sufficiently accurate solution the number of basis functions in the expansion (\ref{PsiExpansion}) should be much larger than unity. The approximations accepted in this work, that all electrons are spin-polarized and occupy only the lowest Landau level states, thus fail at least at $1.0\le\beta\le 1.4211$. To get a more accurate solution of the problem at $\beta\lesssim 1.42$ one should modify the theory by inclusion more many-body states in the expansion (\ref{PsiExpansion}), i.e., by taking into consideration higher Landau level states. 

In intermediate magnetic fields, $1.4211\le \beta< 1.721$, the total angular momentum in the ground state is ${\cal L}=28$, and the density $n_e(r)/n_s$ assumes the form of a broad ring with a single density maximum at a finite $r$ and a minimum at $r=0$. This shape of the electron density resembles the Wigner molecule in the configuration $(0,N)$ and is consistent with the unusual change $\delta{\cal L}=7$ in the ground state ${\cal L}$-sequence which has been noted in Section \ref{sec:prelim-sum}. The state ${\cal L}=27$ with the ``correct'' change $\delta{\cal L}=6$ does have the electron density corresponding to the configuration $(1,6)$, with maxima at $r=0$ and at a finite $r=R_{\max}$, see Figure \ref{fig:Density_N7lowB}(b), but it is not the ground but the first excited state in the system of seven electrons in the considered interval of $\beta$.

What is the reason for such an unusual behavior of the electron density at $1.4211\le \beta< 1.721$? I see two possible explanations here.

First, it is not entirely clear whether the solution in this range of $\beta$ is sufficiently accurate. If ${\cal L}=28$, then the number of many-body basis states in the expansion (\ref{PsiExpansion}) is $N_{mbs}=15$. Although this number is much larger than one, it is worth checking whether this result remains valid when states from higher Landau levels are taken into account. So a more general theory is needed here.

However, if this result is correct, that is, if in moderate magnetic fields $1.4211\le\beta\le 1.7210$ the quantum-mechanically calculated ground state really has a single-shell configuration $(0.7)$, then this can be explained in the following way. 

If the magnetic field is strong, then the length $\lambda\propto 1/\sqrt{\beta}$, which determines the size of the wave functions, is small, and the electrons are quite ``thin'' like classical point particles. Trying to find the configuration with the lowest energy in the field of the attractive potential shown in Figure \ref{fig:Ubpotential}, six electrons push one of their fellows into the center of the disk, and themselves form a ring around it. Thus, they form the configuration $(1,6)$. As the magnetic field decreases, the length $\lambda$ grows, and the electrons become ``thick''. Seven ``fat'' electrons cannot place one of their fellows in the center of the potential well, because it would cost energy. In this case, all seven electrons are located at the edge of the well, creating a single-shell configuration $(0,7)$.

The question of which of the two situations described is the case in reality can be answered after a more complete theory has been developed, involving states from higher Landau levels.

\subsubsection{Seven particles, smooth density profile\label{sec:density7smoo}}

\begin{figure}[ht!]
\includegraphics[width=0.49\columnwidth]{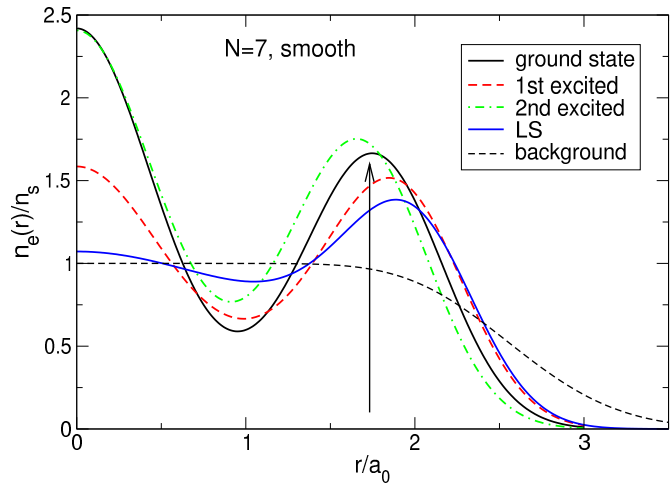}
\caption{\label{fig:densN7smooth} The density of electrons in the ground state (${\cal L}=57$), first (${\cal L}=63$) and second (${\cal L}=51$) excited states, and in the Laughlin state in a system of $N=7$ particles at $\beta=3$. The positive background density profile is smooth. The arrow at $r/a_0=R_s/a_0=1.732047$ shows the position of the outer shell of the classical Wigner molecule in the case of the smooth background density. The density of the Laughlin state at $r=0$ is 2.2576 times smaller than the density of the ground state. }
\end{figure}

In the model of the smooth background density profile, Eq. (\ref{dens-smooth}), the electron density plots do not differ significantly from those for the step-like profile. Therefore I only show one picture with the densities of three exact lowest-energy states (${\cal L}=57$, 63, and 51) and the density of the Laughlin state at $\beta=3$ and $N=7$. The energies of these states are shown in Figure \ref{fig:energyN7smoo}(b) and in Table \ref{tab:N7EnergyGS1st2ndLauSmooth}. Now Figure \ref{fig:densN7smooth} shows the density of electrons in these states. 

One sees that the density of electrons in the ground state has a high maximum in the disk center. The local density here is 2.42 times larger than the density of the positive background. The position of the second maximum $R_{\max}$ is very close to the shell radius in the classical Wigner molecule $R_s$. The densities of the first and second excited states have a similar structure, with the maxima at $r=0$ and near $r=R_s$. This distribution of electrons is physically clear and fully consistent with the intuitive understanding of the physics of Coulomb interacting particles.

In contrast, the density of electrons in the Laughlin state qualitatively differs from those of the ground and the two excited states. In the Laughlin state, the electron density at the center of the disk is very low, it is 2.2576 times less than in the ground state at $r=0$, and the density maximum is located at the point $r/a_0=1.888$, located significantly farther from the center of the disk than the radius of the classical shell in the $(1,6)$ configuration. So, all seven electrons in the Laughlin state accumulate at the edge of the deep potential well, Figure \ref{fig:Ubpotential}, and none of them want to jump to the bottom of the well, although this would obviously lower the energy of the system.

\section{Summary and conclusions\label{sec:conclusion}}

The theory presented in Ref. \cite{Mikhailov23a} and in this paper gives a simple and physically clear picture of the ground and excited states of a system of $N$ 2D electrons at the lowest Landau level. Classical point particles interacting through Coulomb forces form a Wigner crystal molecule, in the configuration $(0,N)$ if $N\lesssim 6$ and $(1,N-1)$ if $N\gtrsim 6$. The quantum-mechanical solution has the form very close to the classical distribution of particles. For $N\le 5$, the density of electrons has the form of a ring with radius $R_{\max}$ close to the radius of the classical Wigner molecule $R_s$. At $N=7$, in addition to the ring with the radius close to $R_s$, a maximum appears at the disk center, reflecting the situation with two shells $(1,N-1)$ in the classical Wigner molecule.

The ring radius $R_{\max}$ in the quantum mechanical solution depends on the magnetic field parameter $\beta$ and the total angular momentum in the ground state ${\cal L}$. As the magnetic field increases, the scale of the wave functions $\lambda$ decreases, which leads to a monotonic decrease in the ring radius $R_{\max}$. At some $\beta$-point, the difference between $R_{\max}$ and $R_s$ becomes too large, and the system jumps to a state with a larger ${\cal L}$. In this way, a discrete increase in angular momentum by $\delta {\cal L}$ compensates for the accumulated monotonic decrease in $R_{\max}$ caused by the growth of $B$. 

As a result, ground states characterized by different values of ${\cal L}$ replace each other as $B$ increases, which leads to the appearance of energy gaps between the ground state and the first excited states of the system. The values of the energy gaps oscillate with $B$, disappearing at certain $\beta$-points and remaining finite between such points. Typical values of the energy gaps are on the order of $0.05 e^2/a_0\epsilon$ (I restore the dielectric permittivity of the environment in this final formula), which corresponds to about $0.3-0.5$ meV under typical experimental conditions. This reasonably agrees with typical energy scales in the FQHE experiments. 

Further development of the theory may include calculations of the Hall conductivity in the alternating ground states and extension of the theory to filling factors $\nu\gtrsim 1$ by including states of higher Landau levels. Calculations for systems with a larger number of particles could also shed additional light on the true nature of the fractional quantum Hall effect.

\appendix

\section{Integral ${\cal J}$\label{app:intJ}}

For the smooth density profile the matrix elements of the background-background, background-electron and electron-electron interactions can be expressed in terms of the integral ${\cal J}$ defined as
\be 
{\cal J}(n_1,n_2,l_1,l_2,k;\alpha,\beta)=\sqrt{\frac 8\pi}\int_0^\infty dx x^{2k} 
L_{n_1}^{l_1}\left(\alpha x^2\right)
L_{n_2}^{l_2}\left(\beta x^2\right)
e^{-(\alpha+\beta)x^2} 
\label{intJ},
\ee
where $n_1$, $n_2$, $l_1$, $l_2$, and $k$ are integers, $\alpha$, $\beta$ are real numbers, and $L_n^l(x)$ are Laguerre polynomials. Substituting the explicit expression for the generalized Laguerre polynomial into the definition (\ref{intJ}) one gets a finite sum of integrals of the type $\int_0^\infty x^{2n}e^{-ax^2}dx$, which can be analytically calculated. The final result can be written in the form
\be 
{\cal J}(n_1,n_2,l_1,l_2,k;\alpha,\beta)=
\sqrt{\frac{2}{\pi(\alpha+\beta)}}
\sum_{m_1=0}^{n_1}\left(\begin{array}{c}
n_1+l_1 \\ m_1+l_1 \\
\end{array}\right)\frac{(-\alpha)^{m_1}}{m_1!}
\sum_{m_2=0}^{n_2}\left(\begin{array}{c}
n_2+l_2 \\ m_2+l_2 \\
\end{array}\right)\frac{(-\beta)^{m_2}}{m_2!}
  \frac{\Gamma\left(m_1+m_2+k+\frac 12\right)}{(\alpha+\beta)^{m_1+m_2+k}}.
\label{intJsolution}
\ee
The integrals ${\cal I}(n_1,n_2,k)$ and ${\cal K}(n_1,n_2,k)$, which have been defined in the previous paper \cite{Mikhailov23a}, are related to ${\cal J}$ as follows
\be 
{\cal I}(n_1,n_2,k)=
{\cal J}\left(n_1,n_2,k,k, k;1,1\right),
\label{intI}
\ee
\be 
{\cal K}(n_1,n_2,k)=\sqrt{\frac{n_1!}{(n_1+k)!}\frac{n_2!}{(n_2+k)!}}{\cal I}(n_1,n_2,k).
\label{intK}
\ee

%\bibliography{../../../BIB-FILES/mikhailov,../../../BIB-FILES/lowD,../../../BIB-FILES/wc,../../../BIB-FILES/fqhe,../../../BIB-FILES/math} 
\bibliography{}

\begin{thebibliography}{12}%
\makeatletter
\providecommand \@ifxundefined [1]{%
 \@ifx{#1\undefined}
}%
\providecommand \@ifnum [1]{%
 \ifnum #1\expandafter \@firstoftwo
 \else \expandafter \@secondoftwo
 \fi
}%
\providecommand \@ifx [1]{%
 \ifx #1\expandafter \@firstoftwo
 \else \expandafter \@secondoftwo
 \fi
}%
\providecommand \natexlab [1]{#1}%
\providecommand \enquote  [1]{``#1''}%
\providecommand \bibnamefont  [1]{#1}%
\providecommand \bibfnamefont [1]{#1}%
\providecommand \citenamefont [1]{#1}%
\providecommand \href@noop [0]{\@secondoftwo}%
\providecommand \href [0]{\begingroup \@sanitize@url \@href}%
\providecommand \@href[1]{\@@startlink{#1}\@@href}%
\providecommand \@@href[1]{\endgroup#1\@@endlink}%
\providecommand \@sanitize@url [0]{\catcode `\\12\catcode `\$12\catcode
  `\&12\catcode `\#12\catcode `\^12\catcode `\_12\catcode `\%12\relax}%
\providecommand \@@startlink[1]{}%
\providecommand \@@endlink[0]{}%
\providecommand \url  [0]{\begingroup\@sanitize@url \@url }%
\providecommand \@url [1]{\endgroup\@href {#1}{\urlprefix }}%
\providecommand \urlprefix  [0]{URL }%
\providecommand \Eprint [0]{\href }%
\providecommand \doibase [0]{https://doi.org/}%
\providecommand \selectlanguage [0]{\@gobble}%
\providecommand \bibinfo  [0]{\@secondoftwo}%
\providecommand \bibfield  [0]{\@secondoftwo}%
\providecommand \translation [1]{[#1]}%
\providecommand \BibitemOpen [0]{}%
\providecommand \bibitemStop [0]{}%
\providecommand \bibitemNoStop [0]{.\EOS\space}%
\providecommand \EOS [0]{\spacefactor3000\relax}%
\providecommand \BibitemShut  [1]{\csname bibitem#1\endcsname}%
\let\auto@bib@innerbib\@empty
%</preamble>
\bibitem [{\citenamefont {Tsui}\ \emph {et~al.}(1982)\citenamefont {Tsui},
  \citenamefont {Stormer},\ and\ \citenamefont {Gossard}}]{Tsui82}%
  \BibitemOpen
  \bibfield  {author} {\bibinfo {author} {\bibfnamefont {D.~C.}\ \bibnamefont
  {Tsui}}, \bibinfo {author} {\bibfnamefont {H.~L.}\ \bibnamefont {Stormer}},\
  and\ \bibinfo {author} {\bibfnamefont {A.~C.}\ \bibnamefont {Gossard}},\
  }\bibfield  {title} {\bibinfo {title} {Two-dimensional magnetotransport in
  the extreme quantum limit},\ }\href@noop {} {\bibfield  {journal} {\bibinfo
  {journal} {Phys. Rev. Lett.}\ }\textbf {\bibinfo {volume} {48}},\ \bibinfo
  {pages} {1559} (\bibinfo {year} {1982})}\BibitemShut {NoStop}%
\bibitem [{\citenamefont {{von Klitzing}}\ \emph {et~al.}(1980)\citenamefont
  {{von Klitzing}}, \citenamefont {Dorda},\ and\ \citenamefont
  {Pepper}}]{Klitzing80}%
  \BibitemOpen
  \bibfield  {author} {\bibinfo {author} {\bibfnamefont {K.}~\bibnamefont {{von
  Klitzing}}}, \bibinfo {author} {\bibfnamefont {G.}~\bibnamefont {Dorda}},\
  and\ \bibinfo {author} {\bibfnamefont {M.}~\bibnamefont {Pepper}},\
  }\bibfield  {title} {\bibinfo {title} {New method for high-accuracy
  determination of the fine-structure constant based on quantized {H}all
  resistance},\ }\href@noop {} {\bibfield  {journal} {\bibinfo  {journal}
  {Phys. Rev. Lett.}\ }\textbf {\bibinfo {volume} {45}},\ \bibinfo {pages}
  {494} (\bibinfo {year} {1980})}\BibitemShut {NoStop}%
\bibitem [{\citenamefont {Landau}(1930)}]{Landau30}%
  \BibitemOpen
  \bibfield  {author} {\bibinfo {author} {\bibfnamefont {L.}~\bibnamefont
  {Landau}},\ }\bibfield  {title} {\bibinfo {title} {Diamagnetismus der
  metalle},\ }\href@noop {} {\bibfield  {journal} {\bibinfo  {journal}
  {Zeitschrift für Physik}\ }\textbf {\bibinfo {volume} {64}},\ \bibinfo
  {pages} {629–637} (\bibinfo {year} {1930})}\BibitemShut {NoStop}%
\bibitem [{\citenamefont {Prange}\ and\ \citenamefont
  {Girvin}(1990)}]{Prange90}%
  \BibitemOpen
  \bibinfo {editor} {\bibfnamefont {R.~E.}\ \bibnamefont {Prange}}\ and\
  \bibinfo {editor} {\bibfnamefont {S.~M.}\ \bibnamefont {Girvin}},\ eds.,\
  \href@noop {} {\emph {\bibinfo {title} {The Quantum Hall Effect}}}\ (\bibinfo
   {publisher} {Springer},\ \bibinfo {address} {New York},\ \bibinfo {year}
  {1990})\BibitemShut {NoStop}%
\bibitem [{\citenamefont {Laughlin}(1983)}]{Laughlin83}%
  \BibitemOpen
  \bibfield  {author} {\bibinfo {author} {\bibfnamefont {R.~B.}\ \bibnamefont
  {Laughlin}},\ }\bibfield  {title} {\bibinfo {title} {Anomalous quantum {H}all
  effect: An incompressible quantum liquid with fractionally charged
  excitations},\ }\href@noop {} {\bibfield  {journal} {\bibinfo  {journal}
  {Phys. Rev. Lett.}\ }\textbf {\bibinfo {volume} {50}},\ \bibinfo {pages}
  {1395} (\bibinfo {year} {1983})}\BibitemShut {NoStop}%
\bibitem [{\citenamefont {Bychkov}\ \emph {et~al.}(1981)\citenamefont
  {Bychkov}, \citenamefont {Iordanskii},\ and\ \citenamefont
  {Eliashberg}}]{Bychkov81}%
  \BibitemOpen
  \bibfield  {author} {\bibinfo {author} {\bibfnamefont {Y.~A.}\ \bibnamefont
  {Bychkov}}, \bibinfo {author} {\bibfnamefont {S.~V.}\ \bibnamefont
  {Iordanskii}},\ and\ \bibinfo {author} {\bibfnamefont {G.~M.}\ \bibnamefont
  {Eliashberg}},\ }\bibfield  {title} {\bibinfo {title} {Two-dimensional
  electrons in a strong magnetic field},\ }\href@noop {} {\bibfield  {journal}
  {\bibinfo  {journal} {JETP Lett.}\ }\textbf {\bibinfo {volume} {33}},\
  \bibinfo {pages} {143} (\bibinfo {year} {1981})}\BibitemShut {NoStop}%
\bibitem [{\citenamefont {Yoshioka}\ and\ \citenamefont
  {Fukuyama}(1979)}]{Yoshioka79}%
  \BibitemOpen
  \bibfield  {author} {\bibinfo {author} {\bibfnamefont {D.}~\bibnamefont
  {Yoshioka}}\ and\ \bibinfo {author} {\bibfnamefont {H.}~\bibnamefont
  {Fukuyama}},\ }\bibfield  {title} {\bibinfo {title} {Charge density wave
  state of two-dimensional electrons in strong magnetic fields},\ }\href@noop
  {} {\bibfield  {journal} {\bibinfo  {journal} {J. Phys. Soc. Japan}\ }\textbf
  {\bibinfo {volume} {47}},\ \bibinfo {pages} {394} (\bibinfo {year}
  {1979})}\BibitemShut {NoStop}%
\bibitem [{\citenamefont {Yoshioka}\ and\ \citenamefont
  {Lee}(1983)}]{Yoshioka83b}%
  \BibitemOpen
  \bibfield  {author} {\bibinfo {author} {\bibfnamefont {D.}~\bibnamefont
  {Yoshioka}}\ and\ \bibinfo {author} {\bibfnamefont {P.~A.}\ \bibnamefont
  {Lee}},\ }\bibfield  {title} {\bibinfo {title} {Ground-state energy of a
  two-dimensional charge-density-wave state in a strong magnetic field},\
  }\href@noop {} {\bibfield  {journal} {\bibinfo  {journal} {Phys. Rev. B}\
  }\textbf {\bibinfo {volume} {27}},\ \bibinfo {pages} {4986} (\bibinfo {year}
  {1983})}\BibitemShut {NoStop}%
\bibitem [{\citenamefont {Willett}\ \emph {et~al.}(1987)\citenamefont
  {Willett}, \citenamefont {Eisenstein}, \citenamefont {Stormer}, \citenamefont
  {Tsui}, \citenamefont {Gossard},\ and\ \citenamefont {English}}]{Willett87}%
  \BibitemOpen
  \bibfield  {author} {\bibinfo {author} {\bibfnamefont {R.}~\bibnamefont
  {Willett}}, \bibinfo {author} {\bibfnamefont {J.~P.}\ \bibnamefont
  {Eisenstein}}, \bibinfo {author} {\bibfnamefont {H.~L.}\ \bibnamefont
  {Stormer}}, \bibinfo {author} {\bibfnamefont {D.~C.}\ \bibnamefont {Tsui}},
  \bibinfo {author} {\bibfnamefont {A.~C.}\ \bibnamefont {Gossard}},\ and\
  \bibinfo {author} {\bibfnamefont {J.~H.}\ \bibnamefont {English}},\
  }\bibfield  {title} {\bibinfo {title} {Observation of an even-denominator
  quantum number in the fractional quantum {H}all effect},\ }\href@noop {}
  {\bibfield  {journal} {\bibinfo  {journal} {Phys. Rev. Lett.}\ }\textbf
  {\bibinfo {volume} {59}},\ \bibinfo {pages} {1776} (\bibinfo {year}
  {1987})}\BibitemShut {NoStop}%
\bibitem [{\citenamefont {Haldane}(1983)}]{Haldane83}%
  \BibitemOpen
  \bibfield  {author} {\bibinfo {author} {\bibfnamefont {F.~D.~M.}\
  \bibnamefont {Haldane}},\ }\href@noop {} {\bibfield  {journal} {\bibinfo
  {journal} {Phys. Rev. Lett.}\ }\textbf {\bibinfo {volume} {51}},\ \bibinfo
  {pages} {605} (\bibinfo {year} {1983})}\BibitemShut {NoStop}%
\bibitem [{\citenamefont {Jain}(1989)}]{Jain89}%
  \BibitemOpen
  \bibfield  {author} {\bibinfo {author} {\bibfnamefont {J.~K.}\ \bibnamefont
  {Jain}},\ }\bibfield  {title} {\bibinfo {title} {Composite-fermion approach
  for the fractional quantum {H}all effect},\ }\href@noop {} {\bibfield
  {journal} {\bibinfo  {journal} {Phys. Rev. Lett.}\ }\textbf {\bibinfo
  {volume} {63}},\ \bibinfo {pages} {199} (\bibinfo {year} {1989})}\BibitemShut
  {NoStop}%
\bibitem [{\citenamefont {Mikhailov}()}]{Mikhailov23a}%
  \BibitemOpen
  \bibfield  {author} {\bibinfo {author} {\bibfnamefont {S.~A.}\ \bibnamefont
  {Mikhailov}},\ }\href@noop {} {\bibinfo {title} {Toward a correct theory of
  the fractional quantum {H}all effect: {W}hat is the ground state of a quantum
  {H}all system at $\nu = 1/3$?}},\ \bibinfo {note} {arXiv:2206.05152v3
  (2023)}\BibitemShut {NoStop}%
\end{thebibliography}%
\end{document}